\documentstyle[12pt]{article}
\catcode`\@=11
\def\newsymbol#1#2#3#4#5{\let\next@\relax%
 \ifnum#2=\@ne\else%
 \ifnum#2=\tw@\let\next@\msyfam@\fi\fi%
 \mathchardef#1="#3\next@#4#5}
\def\mathhexbox@#1#2#3{\relax%
 \ifmmode\mathpalette{}{\m@th\mathchar"#1#2#3}r
 \else\leavevmode\hbox{$\m@th\mathchar"#1#2#3$}\fi}
\def\hexnumber@#1{\ifcase#1 0\or 1\or 2\or 3\or 4\or 5\or 6\or 7\or 8%
\or 9\or A\or B\or C\or D\or E\or F\fi}
\font\tenmsy=msbm10
\font\sevenmsy=msbm7
\font\fivemsy=msbm5
\newfam\msyfam
\textfont\msyfam=\tenmsy
\scriptfont\msyfam=\sevenmsy
\scriptscriptfont\msyfam=\fivemsy
\edef\msyfam@{\hexnumber@\msyfam}
\def\Bbb#1{\fam\msyfam\relax#1}
\catcode`\@=\active
\load{\footnotesize}{\sf}

\newtheorem{theorem}{Theorem}[section]
\newtheorem{proposition}[theorem]{Proposition}
\newtheorem{lemma}[theorem]{Lemma}
\newtheorem{corollary}[theorem]{Corollary}
\newtheorem{definition}[theorem]{Definition}

\newtheorem{remark}[theorem]{Remark}

\newcommand{\proof}{{\noindent \it Proof:\ }}
\newcommand{\qed}{\hfill $\Box$\par\medskip}
\newcommand{\BR}{{{\Bbb R}^3}}
\newcommand{\BC}{{\Bbb C}}
\newcommand{\RR}{{{\Bbb R}}}
\newcommand{\RRR}{\RR^{3n}}
\newcommand{\F}{\prod_{j=1}^n f_j(k_j)} 
\newcommand{\G}[2]{
\lk\!\!\!
\begin{array}{c}#1 \\ #2\end{array}
\!\!\!\rk }

\newcommand{\kr}{(\kn)\in \RR^{3n}} 
 
\newcommand{\kn}{k_1,...,k_n} 
\newcommand{\knl}{k_1,...,k_l} 
\newcommand{\mpp}{-\half\Delta} 
\newcommand{\knn}{\k_1,...,\k_n}

\newcommand{\LR}{{L^2({\Bbb R}^3)}}
\newcommand{\LRr}{{L_{\rm real}^2({\Bbb R}^3)}}
\newcommand{\LRN}{{L^2({\Bbb R}^3\times\jjj)}}
\newcommand{\LRn}{{L^2({\Bbb R}^4)}} 
\newcommand{\x}{x,X_\cdot} 
\newcommand{\LRnr}{{L_{\rm real}^2({\Bbb R}^4)}} 
\newcommand{\vvv}{e^{-\int_0^t V(X_s) ds}} 
\newcommand{\vvvv}{e^{-2\int_0^t V(X_s) ds}} 
\newcommand{\vvvvv}{e^{-4\int_0^t V(X_s) ds}} 
\newcommand{\J}{{\cal J}_t}
\newcommand{\EEE}{{\Bbb E}_x} 
\newcommand{\EEEq}{{\Bbb E}^Q_x} 

\newcommand{\ccccc}{C_0^\infty(\BR\setminus\{0\})} 
\newcommand{\cccc}[1]{c_#1\|\KK_{m_#1}\Psi\|_\hhh}
\newcommand{\Psim}{\{\Psi_m\}} 
\newcommand{\Psid}{\Psi_{m'}} 
\newcommand{\mdd}{\{m'\}} 
\newcommand{\md}{\{m\}} 
\newcommand{\ch}{C^\infty(\hf)} 
\newcommand{\lll}{l-\sum_{i=1}^{n-1}p_i} 
\newcommand{\phis}{\phi_\ast} 
\newcommand{\hm}{{\Bbb H}_m} 
\newcommand{\cx}{C^\infty(|x|)} 
\newcommand{\KK}{K}
\newcommand{\PPP}{\lk    \left\|\frac{e\vp}{\sqrt{2\omega}}\right\|^2\rk}
\renewcommand{\c}{c_1}
\newcommand{\cdott}{\!\cdot\!} 

\newcommand{\cc}{c_2} 

\newcommand{\qqq}{{L^2(Q)}} 
\newcommand{\qqqq}{{L^2(Q_0)}} 
\newcommand{\wick}[1]{:\!\! #1\!\!:} 
\newcommand{\od}{\oplus^3}
\newcommand{\odd}{\oplus_{\mu=1,2,3}}

\newcommand{\jjj}{\{1,2\}}
\newcommand{\jii}{\sum_{j=0,1,2}}
\newcommand{\xxx}{ \sum_{l=0}^n\sum_{\{p_1,..,p_l\}\subset\{1,..,n\}}}

\newcommand{\XXX}{\sum_{k=1}^{n-l} \alk \|\N|x|^{m+l}\Psi\|_\hhh^2}
\newcommand{\alk}{a_k^{n,l}} 
\newcommand{\clk}{c_k^{n,l}} 
\newcommand{\YYY}{\sum_{l=0}^n\sum_{k=1}^{n-l} \clk \|\N |x|^{m+l}\Psi\|_\hhh^2}
\newcommand{\dln}{d^n_{l}}
\newcommand{\ZZZ}{\sum_{l=1}^n \dln} 
\newcommand{\II}{{\cal I}}
\renewcommand{\t}{\theta} 
\newcommand{\R}{{\cal R}} 
\newcommand{\tl}{\t_{p_1}...\t_{p_l}}
\newcommand{\te}{\frac{|e|}{\sqrt 2}}

\newcommand{\tll}{
\lk \prod_{i=1}^l \int  \frac{|e\vp(k)|}{\sqrt{2\omega(k)}} f_{p_i}(k) dk \rk}

\newcommand{\supx}{\sup_{x\in\BR}}

\newcommand{\kj}[1]{#1(\k_1)...#1(\k_n)} 
\newcommand{\kl}[1]{#1(\k_1)...#1(\k_l)} 
\newcommand{\kkj}[1]{#1(\k_1)...\widehat{#1(\k_p)}...#1(\k_n)} 
\newcommand{\kkkj}[1]{#1(\k_1)...\widehat{#1(\k_q)}...\widehat{#1(\k_p)}...#1(\k_n)} 

\newcommand{\kjj}[1]{#1(\k_1)...\widehat{#1 (\k_{p_1})}...\widehat{#1 (\k_{p_l})}...
#1(\k_n)} 
\newcommand{\bln}{b_l^n} 
\newcommand{\limp}{\lim_{p\rightarrow \infty}}
\newcommand{\limn}{\lim_{n\rightarrow \infty}}
\newcommand{\limm}{\lim_{m\rightarrow \infty}}
\newcommand{\limd}{\lim_{m'\rightarrow \infty}}
\newcommand{\jk }[1]{#1^\ast (\k_n)...#1^\ast(\k_1)} 

\newcommand{\dk}{dk_1...dk_n} 
\newcommand{\dkl}{dk_1...dk_l} 
\newcommand{\dkk}{dk_1..\widehat{dk}_{p_1}..\widehat{dk}_{p_l}..dk_n}

\newcommand{\dm}{dk'_1...dk'_l} 
\newcommand{\k}{{\bf k}}
\newcommand{\m}{{\bf k}'}
\newcommand{\M}[1]{{\bf N}_{#1}(\Psi)} 
\newcommand{\nd}{{\cal N}_{\d}(\Psi)} 
\newcommand{\nnd}{{{\cal N}}(\Psi)}

\newcommand{\g}{{{\cal N}}(\gr)}

\newcommand{\ji}{\int \!\!\!\!\!\!\! \sum}
\newcommand{\jji}{\sum_{j=1,2} \int}

\newcommand{\fff}{{\cal F}} 
\newcommand{\fffn}{{\cal F}_{\rm N}} 
\newcommand{\fffw}{{\cal F}_\omega} 
\newcommand{\hhh}{{\cal H}}
\newcommand{\hhnn}{{{\cal H}_n}}
\newcommand{\hhhn}{{{\cal H}_{\rm N}}}
\newcommand{\hn}{H_{\rm N}}
\newcommand{\HH}{\vartheta_1} 
\newcommand{\HHH}{\vartheta_2} 

\newcommand{\f}{^{-1}}

\newcommand{\e}{\epsilon} 
\newcommand{\ek}{\e(\k_1,...,\k_n)}

\newcommand{\sw}{\sump\omega(k_p)} 

\renewcommand{\d}{{\cal D}} 
\newcommand{\E}{{\cal E}} 
\newcommand{\EE}{{\cal C}} 
\newcommand{\ee}{e^{tE}}
\newcommand{\eht}{e^{-tH}}
\newcommand{\ele}{H_{\rm p}}

\newcommand{\ehtp}{e^{-tH_{\rm p}}}
\newcommand{\N}{N^{k/2}}
\newcommand{\NN}{N_0^{k/2}}
\newcommand{\K}{\xi}
\newcommand{\KKK}[1]{(\hff^{#1}+|x|^{2#1})} 
\newcommand{\ov}[1]{\overline{#1}} 

\newcommand{\pk}{P_k(\K)} 
\newcommand{\n}{N_0} 

\newcommand{\ccc}{C_0^\infty(\BR)} 

\newcommand{\mn}{\sum_{\mu,\nu=1,2,3}}
\newcommand{\sump}{\sum_{p=1}^n}

\newcommand{\add}{a^{\ast}}
\newcommand{\ad}{{\rm ad}}
\newcommand{\ass}{a^{\sharp}}
\newcommand{\han}{{1/2}}
\newcommand{\half}{\frac{1}{2}}
\newcommand{\sh}{\frac{1}{\sqrt{2}}}
\newcommand{\HP}{{\cal R}_\omega}

\newcommand{\lk}{\left(}

\newcommand{\rk}{\right)}
\newcommand{\lkk}{\left\{}
\newcommand{\rkk}{\right\}}

\newcommand{\hf}{H_{\rm f}}
\newcommand{\hff}{(H_{\rm f}+1)}

\newcommand{\hfn}{{H_{\rm f}^{\rm N}}}
\newcommand{\hhf}{{\widetilde{H}_{\rm f}}}

\newcommand{\ej}{e(k,j)}

\renewcommand{\S}[1]{{\cal R}_#1} 

\newcommand{\is}{\inf\!\sigma\!}

\newcommand{\vp}{{\widehat{\varphi}}}
\newcommand{\rx}{{\cal R}_x(\k_p)} 
\newcommand{\rrx}{{\cal R}_x(\k_p,\k_q)}

\newcommand{\kx}{e^{ik\cdot x}}
\newcommand{\kxx}{e^{-ik\cdot x}}

\newcommand{\ea}{e^{-ix\cdot A}}
\newcommand{\eaa}{e^{ix\cdot A}}

\newcommand{\eq}[1]{\begin{equation}
\label{#1}}
\newcommand{\en}{\end{equation}}
\newcommand{\bl}[1]{\begin{lemma}
\label{#1}}
\newcommand{\el}{\end{lemma}}

\newcommand{\bt}[1]{\begin{theorem}
\label{#1}}
\newcommand{\et}{\end{theorem}}

\newcommand{\bi}{\begin{description}}
\newcommand{\ei}{\end{description}}

\newcommand{\bp}[1]{\begin{proposition}
\label{#1}}
\newcommand{\ep}{\end{proposition}}

\newcommand{\bc}[1]{\begin{corollary}
\label{#1}}
\newcommand{\ec}{\end{corollary}}

\newcommand{\kak}[1]{(\ref{#1})}

\newcommand{\gr}{\psi_{g}}
\newcommand{\fg}{f_{\rm p}}
\newcommand{\grn}{\psi_{g}^{\rm N}}

\makeatletter
\@addtoreset{equation}{section}
\makeatother

\title
{Localization  of 
the number of photons of ground states in nonrelativistic QED} 
\author{
Fumio Hiroshima\thanks{
Department of Mathematics and Physics,  
Setsunan University,   572-8508, Osaka, Japan, e-mail hiroshima$@$mpg.setsunan.ac.jp}} 
\date{\today}

\begin{document}

\setlength{\baselineskip}{18pt}
\maketitle
\begin{abstract}
One  electron system minimally coupled to a quantized radiation field 
 is considered. 
It is assumed that the quantized  radiation  field is  {\it massless},  and 
{\it no} infrared  cutoff is  imposed. 
The Hamiltonian,  $H$,  of this system 
is defined as a self-adjoint operator acting on 
$\LR\otimes\fff\cong L^2(\BR;\fff)$, 
where $\fff$ is the  Boson Fock space over $L^2(\BR\times\{1,2\})$.  
It is shown that 
the  ground state, $\gr$,  of $H$ belongs to   
$\cap_{k=1}^\infty D(1\otimes N^k)$, 
where $N$ denotes the number operator of $\fff$.  
Moreover it is shown that, 
 for almost every  electron position variable $x\in\BR$ and for  arbitrary $k\geq 0$, 
$\|(1\otimes \N)\gr (x) \|_\fff \leq 
D_ke^{-\delta |x|^{m+1}}$ with  some constants $m\geq 0$,  $D_k>0$,  and $\delta>0$ 
independent of $k$.  
In particular  $\gr\in \cap_{k=1}^\infty 
D (e^{\beta |x|^{m+1}}\otimes N^k)$ for 
$0<\beta<\delta/2$ is obtained. 
\end{abstract}

\section{Introduction}
\subsection{The Pauli-Fierz Hamiltonian}
In this paper one spinless electron  minimally coupled to 
a  massless quantized radiation field is considered. 
It is the so-called  Pauli-Fierz model of 
the nonrelativistic QED. 
The Hilbert space of state vectors of the system is given by 
$$\hhh=\LR\otimes \fff,$$
where $\fff$ denotes the Boson Fock space defined by 
$$\fff=\bigoplus_{n=0}^\infty \left[ 
\otimes_s^n \LRN\right],$$
where 
$\otimes_s^n \LRN$, $n\geq 1$, 
 denotes the $n$-fold symmetric tensor product of $\LRN$ 
and 
$\otimes_s^0\LRN=\BC$. 
The Fock vacuum $\Omega$ is defined by 
$\Omega=\{1,0,0,...\}$. 
Let 
$$\fff_0=\{\oplus_{n=0}^\infty\Psi^{(n)} \in\fff| \Psi^{(n)}=0 \mbox{ for } n\geq m \mbox{ with some } m\}.$$
For each $\{k,j\}\in \BR\times\jjj$, the annihilation operator 
$a (k,j)$ is defined by,  for $\Psi=\oplus_{n=0}^\infty\Psi^{(n)}\in\fff_0$,  
$$\lk a(k,j)\Psi\rk^{(n)}(k_1,j_1...,k_n,j_n)=
\sqrt{n+1}\Psi^{(n+1)}(k,j,k_1,j_1,...,k_n,j_n).$$
The creation operator $\add(k,j)$ is given by 
$\add(k,j)=\lk a(k,j)\lceil_{\fff_0}\rk ^\ast$. 
They satisfy the canonical commutation  relations on $\fff_0$ 
$$[a(k,j), \add(k',j')]=\delta(k-k')\delta_{jj'},$$
$$[a(k,j), a(k',j')]=0 ,$$
$$[\add (k,j), \add(k',j')]=0 .$$
The closed extensions  of $a(k,j)$ and $\add(k,j)$ 
are denoted by  the same symbols respectively. 
The  annihilation and creation operators smeared by $f\in\LR$ are formally written as 
$$\ass(f,j)=\int \ass (k,j)f(k) dk,\ \ \ \ass=a\mbox{ or } \add,$$
and  act as 
$$\lk a(f,j)\Psi\rk^{(n)}=\sqrt{n+1}\int f(k) \Psi^{(n+1)} (k,j,k_1,j_1...,k_n,j_n) dk,$$
$$\lk \add(f,j)\Psi\rk^{(n)}=\frac{1}{\sqrt n} 
\sum_{j_l=j} 
f(k) \Psi^{(n-1)} 
(k_1,j_1,...,\widehat{k_l,j_l}, ..., k_n,j_n) ,$$
where $\sum_{j_l=j}$ denotes to sum up $j_l$ 
such that $j_l=j$, and 
 $\widehat{X}$ means  neglecting  $X$. 
We work with the unit $\hbar=1=c$. 
The dispersion relation is given by 
$$\omega(k)=|k|.$$
Then the free Hamiltonian $\hf$ of $\fff$ is formally written as 
$$\hf =\jji \omega(k) \add(k,j) a(k,j) dk,$$
and acts as 
$$(\hf \Psi)^{(n)}(k_1,j_1,...,k_n,j_n)=\sum_{j=1}^n \omega(k_j)
\Psi^{(n)}(k_1,j_1,...,k_n,j_n),\ \ \ n\geq 1,$$
$$(\hf \Psi)^{(0)}=0$$
with the domain 
$$D(\hf)=\lkk\left.\Psi=\oplus_{n=0}^\infty\Psi^{(n)}\right|\sum_{n=0}^\infty 
\|(\hf \Psi)^{(n)}\|_{\otimes^nL^2(\BR\times\{1,2\})}^2<\infty\rkk.$$
Since $\hf$ is essentially self-adjoint and nonnegative, 
we denotes the self-adjoint extension of $\hf$ by the same symbol $\hf$.  
 Under the identification 
$$\hhh\cong \int_\BR^\oplus \fff dx,$$
the   quantized radiation   field $A$ with a form factor $\varphi$ is given by 
the constant fiber direct integral 
$$A=\int_\BR^\oplus A(x) dx,$$ 
where $A(x)$ is the  operator acting on $\fff$ defined by 
$$A(x)=\sh\jji \frac{\ej}{\sqrt{\omega(k)}} 
 \lkk \add(k,j)\kxx {\vp(-k)}+ 
a (k,j)\kx {\vp(k)}\rkk dk.$$
Here $\vp$ denotes the Fourier transform of $\varphi$ and 
$\ej$, $j=1,2$, are  polarization vectors such that $(e(k,1), e(k,2), k/|k|)$ forms 
a right-handed system, i.e., $k\cdot e(k,j)=0$, $e(k,j)\cdot e(k,j')=\delta_{jj'}$, 
and $e(k,1)\times e(k,2)=k/|k|$ for almost every $k\in\BR$. 
We fix polarization vectors through  this paper. 

The decoupled Hamiltonian is given by 
$$H_0=H_{\rm p}\otimes 1+1\otimes \hf.$$ 
Here 
$$H_{\rm p}=\half p^2+V$$
denotes a particle Hamiltonian, 
where  
$p=(-i\nabla_{x_1}, -i\nabla_{x_2}, -i\nabla_{x_3})$ and $x=(x_1,x_2,x_3)$ 
are   
the momentum operator  and its conjugate position operator 
in $\LR$, respectively, 
and 
$V:\BR\rightarrow \RR$  an external potential. 
We are prepared to define the 
total Hamiltonian, $H$,  of this system, which  is give by 
the minimal coupling to $H_0$. I.e., we replace $p\otimes 1$ with $p\otimes 1-e A$,  
$$H=\half (p\otimes 1-eA)^2+V\otimes  1+1\otimes \hf,$$
where  $e$ denotes 
the   charge of an electron.

\subsection{Assumptions on $V$ and fundamental facts}

We give  assumptions on external potentials. 
We say $V\in K_3$ (the three dimensional Kato class \cite{si}) if 
and only if 
$$\lim_{\epsilon\downarrow 0} 
 \supx \int _{|x-y|<\epsilon} \frac{|V(y)|}{|x-y|} dy =0,$$
and $V\in K_3^{\rm loc}$ if and only if $1_RV\in K_3$ for all $R\geq 0$, 
where 
$$1_R(x)=\lkk\begin{array}{cc}\!\!1,&\! |x|<R,\\
\!\!0,&\!|x|\geq R.\end{array}
\right.$$
Let us define classes $K$ and $V_{\rm exp}$ as follows. 
\begin{definition}
\label{1}
\bi
\item[(1)] We say $V\in K$ if and only if  
$V=V_+-V_-$ such that $V_\pm\geq 0$, 
$V_+\in K^{\rm loc}_3$  and $V_-\in K_3$. 
\item[(2)] We say $V\in V_{\rm exp}$ if and only if 
$V=Z+W$ such that $\inf Z>-\infty$, $Z\in  L_{\rm loc}^1(\BR)$, 
$W<0$, and $W\in L^p(\BR)$ for some $p>3/2$.
\end{description}
\end{definition}
For $V\in K$ a functional integral representation of $e^{-t(\mpp+V)}$ 
by means of the  Wiener measure on $C([0,\infty);\BR)$ is obtained. 
See e.g.,\cite{si}. 
For $V\in K\cap V_{\rm exp}$, using this functional integral representation,  
it can be proven  that a ground state, $\fg$, of $\mpp+V$ 
decays exponentially, i.e., 
\eq{decay}
|\fg(x)|\leq c_1 e^{-c_2|x|^{c_3}}
\en 
for almost every $x\in\BR$ with some positive constants $c_1,c_2,c_3$.  
Similar estimates are available to the Pauli-Fierz Hamiltonian 
$H$ with $V\in K\cap V_{\rm exp}$. 
See Proposition \ref{p1}. 
Furthermore we need to define class 
$V(m), m=0,1,2,...$ to  estimate  constant  $c_3$ in \kak{decay} 
precisely.
\begin{definition}
\label{2}
Suppose that $V=Z+W\in V_{\rm exp}\cap  K$, 
where the decomposition $Z+W$ 
is that of the definition of $V_{\rm exp}$. 
\bi
\item[(1)] We say $V\in V(m)$, $m\geq1$, 
if and only if $Z(x)\geq \gamma |x|^{2m}$ for 
$x\not\in {\cal O}$ with a certain  compact set ${\cal O}$ 
and with  some   $\gamma>0$.
\item[(2)] We say $V\in V(0)$ if and only if 
$\liminf_{|x|\rightarrow \infty}Z(x)> \inf\sigma(H)$, 
where $\sigma(H)$ denotes the spectrum of $H$.
\ei
\end{definition}
A physically reasonable example of $V$ is 
the Coulomb potential 
$\displaystyle \frac{-e Z}{4\pi |x|}$,  
where 
$Z>0$ denotes the charge of a nucleus.  
Actually we see the following proposition.
\bp{ex}
Assume that 
$$  \int_\BR \frac{|\vp(k)|^2}{\omega(k)} dk < 
\frac{Z^2}{2(4\pi)^2}.$$
Then 
$$-\frac{e Z}{4\pi|x|}\in V(0)$$ for all $e>0$.
\ep
\proof 
It is known that $-1/|x|\in K_3\cap V_{\rm exp}$. 
Then we shall show $\is (H)<0$. 
Let $V=-e Z/(4\pi|x|)$ and 
$f$ be the ground state of $\ele=\mpp+V$, $\ele f=-E_0 f$, 
where 
$$E_0=\frac{e^2Z^2}{2(4\pi)^2}.$$ 
Then we have 
$$\is(H) \leq (f\otimes \Omega, Hf\otimes \Omega)_\hhh=
(f,\ele f)_\LR+\frac{e^2}{2}
(f\otimes \Omega, A^2 f\otimes \Omega)_\hhh$$
$$=-E_0+\frac{e^2}{2} 
\sum _{\mu=1,2,3} 
\int_\BR\lk 1-\frac{k_\mu^2}{|k|^2}\rk
\frac{|\vp(k)|^2}{\omega(k)} dk=-\frac{e^2}{2}\lk 
\frac{Z^2}{(4\pi)^2}-
2\int_\BR\frac{|\vp(k)|^2}{\omega(k)} dk
\rk<0.$$
Thus the proposition follows.
\qed

We introduce Hypothesis $\hm$, $m=0,1,2,...$.  \\
{\bf Hypothesis $\hm$ }
{\it
\bi
\item[(1)] $D(\Delta)\subset D(V)$ and 
there exists $0\leq a <1$ and $0\leq b$ such that for $f\in D(\Delta)$, 
$$\|V f\|_\LR\leq a\|\Delta f\|_\LR+b\|f\|_\LR,$$
\item[(2)] $\vp(-k)=\overline{\vp(k)}$, and $\vp/\omega, \sqrt\omega\vp\in\LR$, 
\item[(3)] $\inf\sigma _{\rm ess}(H_{\rm p})-\inf\sigma(H_{\rm p})>0$, 
where $\sigma(H_{\rm p})$ (resp. $\sigma_{\rm ess}(H_{\rm p})$) 
 denotes the spectrum (resp. essential spectrum)  of $H_{\rm p}$, 
\item[(4)] $V\in V(m)$.
\ei
}
\bp{hiroshima}
We assume (1) and (2) of $\hm$. 
Then for arbitrary $e\in\RR$, 
$H$ is self-adjoint on 
$D(\Delta\otimes 1)\cap D(1\otimes \hf)$ and bounded from below, 
moreover essentially self-adjoint on any core of 
$-\Delta\otimes 1 +1\otimes \hf$. 
\ep
\proof
See  \cite{h11, h16}. 
\qed
The number operator of $\fff$ is defined by 
$$N=\jji  \add(k,j) a(k,j) dk.$$ 
The operator $N^k$, $k\geq 0$,   
acts as,  for $\Psi=\oplus_{n=0}^\infty \Psi^{(n)}$,  
$$(N^k\Psi)^{(n)}=n^k\Psi^{(n)}$$
with the domain 
$$D(N^k)=\lkk\left. \Psi=\oplus_{n=0}^\infty \Psi^{(n)} \right | 
\sum_{n=0}^\infty n^{2k}\|\Psi^{(n)}\|_{\otimes^nL^2(\BR\times\{1,2\})}^2<\infty\rkk.$$
We give a remark on notations. 
We can identify $\hhh$ with the set of $\fff$-valued $L^2$-functions on $\BR$, 
i.e., 
\eq{ident}
\hhh\cong L^2(\BR;\fff).
\en 
Under this identification, $\Psi\in \hhh$ can be regarded  as 
a vector in $L^2(\BR;\fff)$. 
Namely  for almost every $x\in\BR$, 
$$\Psi(x)\in \fff.$$
We use identification \kak{ident} without notices in what follows. 
The following proposition is well known. 
\bp{p1}
Suppose $\hm$. Then there exists $e_0\leq\infty$ such that for all $|e|\leq e_0$,  
(i) $H$ has a  ground state $\gr$, (ii) it is unique,  
(iii) $\|(1\otimes N^\han) \gr\|_\hhh<\infty$, 
(iv) $\|\gr(x)\|_\fff\leq D e^{-\delta |x|^{m+1}}$ 
for almost every $x\in\BR$ with some constants $D>0$ and 
$\delta>0$. 
\ep
\proof See \cite{bfs3,gll} for (i) and (iii), 
\cite{h8} for (ii) and \cite{h20} for (iv). \qed
\begin{remark}
It is {\it not} clear directly from  Proposition \ref{p1} that 
$\gr\in D(e^{\delta |x|^{m+1}}\otimes N^\han)$. 
See Corollary \ref{15}. 
\end{remark}
The condition 
\eq{in}
{\cal I}=\int_\BR \frac{|\vp(k)|^2}{\omega(k)^3} dk<\infty 
\en 
is  called the infrared cutoff condition. 
\kak{in} is {\it not} assumed in Proposition 
\ref{p1}. 
For suitable external potentials,  
$e_0=\infty$ is available in Proposition \ref{p1}. 
This is established in \cite{gll}. 
In the case where 
$\inf_{\rm ess}(H_{\rm p})-\inf\sigma(H_{\rm p})=0$, 
examples for $H$ to have  a ground state is investigated  in \cite{hisp1,hvv}. 
It is unknown, however, whether  such a ground state decays in $x$ 
exponentially or not.  
When electron includes spin, 
$H$ has a twofold degenerate ground state for sufficiently small $|e|$, 
which is shown in  \cite{hisp2}.

\subsection{Localization of the number of bosons 
and infrared singularities for a linear coupling model}
The  Nelson Hamiltonian \cite{ne} 
describes  a linear coupling between  a nonrelativistic 
particle and a scalar quantum field with a form factor $\varphi$. 
Let $\hhhn=\LR\otimes \fffn$, where 
$\fffn=\bigoplus_{n=0}^\infty[\otimes_n^s \LR]$. 
The Nelson Hamiltonian is defined as 
a self-adjoint operator acting in the  Hilbert space $\hhhn$, 
which is  
given by 
$$\hn=\ele\otimes 1+1\otimes \hfn+g\phi,$$
where $g$ denotes a coupling constant, 
$\hfn=\int\omega(k)\add(k) a(k) dk$ is the free Hamiltonian in $\fffn$,  
and under identification $\hhhn\cong \int_\BR^\oplus 
 \fffn dx$, $\phi$ is defined by 
$\phi=\int_\BR^\oplus \phi(x) dx$ with 
$$\phi(x)=\sh \int \lkk 
\add(k) e^{-ikx}\frac{\vp(k)}{\sqrt{\omega(k)}} +
a(k) e^{ikx}\frac{\vp(k)}{\sqrt{\omega(k)}} \rkk dk.$$ 
It has been established in \cite{ah, bfs2, ge, sp} 
that the Nelson Hamiltonian has the  unique ground state, $\grn$, 
under the condition 
$$\II<\infty.$$ 
Let us denote the number operator of $\fffn$ 
by the same symbol $N$ as that of $\fff$. 
In \cite{bhlms} it has been  proven that
$\grn$ decays  superexponentially, i.e., 
\eq{3}
\|e^{+\beta (1\otimes N)}\grn\|_\hhhn <\infty
\en 
for arbitrary $\beta>0$. 
This kind of results  has been obtained in  \cite[Section 3]{gr}  and \cite{sl} for 
relativistic polaron models, and \cite[Section 8]{sp3} for  spin-boson models. 
Moreover in \cite{bhlms} we see that 
\eq{pk}
\lim_{{\cal I}\rightarrow\infty}\|(1\otimes N^\han) \grn\|_\hhhn=\infty.
\en 
Actually in the  infrared divergence case, 
\eq{div}
\II=\infty, 
\en 
it is shown in \cite{lms} that the Nelson Hamiltonian  with some confining external 
potentials 
has no ground states in 
$\hhhn$. Then we have to take a non-Fock representation
to investigate a ground state with \kak{div}. 
See \cite{a,ahh, lms2} for details. 
That is to say, 
as the infrared cutoff is removed, 
the number of bosons of  $\grn$ diverges and the ground state disappears. 
A method to show \kak{3} and \kak{pk} is based on a path integral representation of 
$(\grn, e^{+\beta (1\otimes N)}\grn)_\hhhn$. Precisely it can be  shown that in the case ${\cal I}<\infty$  
there exists a probability measure $\mu$ on $C(\RR;\BR)$ such that for arbitrary $\beta>0$,  
\eq{16}
(\grn, e^{+\beta (1\otimes N)}\grn)_\hhhn =\int _{C(\RR;\BR)} e^{-(g^2/2) 
(1-e^{+\beta})\int_{-\infty}^0ds\int_0^\infty dt 
W(q_s-q_t,s-t)}\mu(dq), 
\en 
where $(q_t)_{-\infty<t<\infty} \in C(\RR;\BR)$,  
 and 
\eq{W}
W(X, T)=
\int_\BR e^{-|T|\omega(k)}
e^{ik\cdot X}\frac{|\vp(k)|^2}{{\omega(k)}} dk.
\en 
Note that the double integral 
$\int_{-T}^0 ds\int_0^T dt 
W(q_s-q_t,s-t)$ is estimated uniformly in path and $T$ as 
\eq{ini}
\left|
\int_{-T}^0ds\int_0^T dt 
W(q_s-q_t,s-t)\right| \leq \II.
\en 
This uniform bound is a core of 
the proof of identity \kak{16}.

\subsection{The main theorems}
In contrast to the Nelson Hamiltonian, 
for the Pauli-Fierz Hamiltonian,  
as is seen in Proposition \ref{p1},   
it is shown that the ground state, $\gr$, exists and 
 $\|(1\otimes N^\han)\gr\|_\hhh<\infty$  even in  the case 
$\II =\infty$. 
We may say 
that the infrared singularity for the Pauli-Fierz Hamiltonian 
is not so singular in comparison with the Nelson Hamiltonian, 
and one may expect that  
\eq{13}
\|e^{+\beta (1\otimes N)}\gr\|_\hhh<\infty
\en  holds for some $\beta>0$ 
under $\II=\infty$. 
 Unfortunately, however,  we can not show \kak{13}, 
since the similar path  integral method as the Nelson Hamiltonian is not available 
on account of the appearance of the so-called  
{\it double  stochastic integral}  (\cite{h8}) 
instead of 
$\int_{-\infty}^0 ds \int_0^\infty  dt W(q_s-q_t, s-t)$ in \kak{16}. 
The double  stochastic integral is formally written as 
\eq{st}
\sum_{\mu,\nu=1,2,3} 
\int_{-\infty}^0 dq_{\mu, s} \int_0^\infty  dq_{\nu,t} W_{\mu\nu}(q_s-q_t, s-t),
\en 
where $(q_s)_{-\infty<s<\infty}=(q_{1,s}, q_{2,s}, q_{3,s})_{-\infty<s<\infty}\in C(\RR, \BR)$ and 
$$W_{\mu\nu}(X, T)=
\int_\BR \lk\delta_{\mu\nu}-\frac{k_\mu k_\nu}{|k|^2}\rk e^{-|T|\omega(k)}
e^{ik\cdot X}\frac{|\vp(k)|^2}{{\omega(k)}} dk.$$
Actually we can not estimate \kak{st} uniformly in path such as \kak{ini}. 
Therefore we are not concerned here with \kak{13}. 
In place of this  we will show  the following theorems. 
\bt{t1}
Assume $\hm$. 
Then 
$ \gr\in \bigcap_{k=1}^\infty D(1\otimes \N).$
\et
\begin{remark}
Theorem \ref{t1}   automatically follows  if one assumes that 
photons have artificial  positive mass, $\nu$, i.e., $\omega(k)=\sqrt{|k|^2+\nu^2}$. 
\end{remark}
\bt{t2}
Assume $\hm$. 
Then for a fixed $k\geq 0$ there exist positive constants 
$D_k$,  and $\delta$ independent of $k$ such that 
\eq{t11}
\|(1\otimes \N) \gr(x)\|_\fff\leq D_ke^{-\delta |x|^{m+1}}
\en 
for almost every $x\in\BR$. 
\et
\begin{remark}
We do {\it not} assume $\II<\infty$ in Theorems \ref{t1} and \ref{t2}.
\end{remark}
From Theorems \ref{t1} and \ref{t2} the following corollary is immediate. 
\bc{15}
Assume $\hm$. Then 
$
\gr\in \bigcap_{k=0}^\infty D(e^{\beta|x|^{m+1}}\otimes \N)
$ 
for $\beta<\delta/2$. 
\ec 
\proof 
Since $\gr\in D(e^{2\beta |x|^{m+1}}\otimes 1) \cap D(1\otimes \N)$ for all $k\geq 0$, 
the corollary follows from the fact that 
$D(e^{2\beta |x|^{m+1}}\otimes 1)\cap D(1\otimes N^k)\subset D(e^{\beta|x|^{m+1}}\otimes \N)$. 
\qed

\subsection{Outline of proofs of the main theorems}
For notational convenience,  
in the following we mostly omit the tensor notation 
$\otimes$, e.g., we express as 
$\hf$ for $1\otimes\hf$, $\ass(k,j)$ for $1\otimes \ass(k,j)$, 
$\Delta$ for $\Delta\otimes 1$, $|x|$ for $|x|\otimes 1$, etc.,  
and set 
$$\k=(k,j)\in \BR\times \jjj$$ and 
$$\ji...\dk=\sum_{j_1,...,j_n=1,2}\int ...\dk.$$ 
The strategy of this paper is as follows. 
We check  in Lemma \ref{b3} that 
\eq{as}
\ji\|\kl a\Psi\|^2_\hhh \dkl<\infty,\ \ \ l=1,2,...,k, 
\en 
if and only if 
$$\Psi\in D(\N).$$
Thus in order to 
 prove  Theorem \ref{t1} 
it is enough to show that 
$\gr\in D(\kl a)$ for almost every 
$(\knl)\in \RR^{3l}$, and 
\eq{asa}
\ji\|\kl a \gr\|^2_\hhh \dkl<\infty
\en 
holds 
for {\it all} $l\geq 0$. 
One subtlety to show \ref{asa} is that we do not assume  $\II<\infty$.  
Bach-Fr\"ohlich-Sigal \cite{bfs3} proved 
\kak{asa}  for  $l=1$. 
We extend it   to  $l\geq 1$.

To see \kak{asa} for all $l$ we make a detour through 
the  modified annihilation operator defined by 
$$b(k,j)=a(k,j)-i\frac{e}{\sqrt 2} 
(x\cdot e(k,j)) 
\frac{e^{-ik\cdot x}}{\sqrt{\omega(k)}}\vp(k).$$
For some $\Psi\in \hhh$ we establish  in Lemma \ref{321}
that 
$$\|\kj a\Psi\|_\hhh$$ 
\eq{R2}
\leq 
\xxx \prod_{j=1}^l \frac{|e\vp(k_{p_j})|}{\sqrt{2\omega(k_{p_j})}}
\|\kjj b |x|^l\Psi\|_\hhh,
\en 
where $\widehat{\ }$ means neglecting  the term below, and 
$\sum_{\{p_1,...,p_l\}\subset\{1,2,...,n\}}$ 
denotes  to sum up  all the combinations to choose $l$ numbers from $\{1,2,...,n\}$. 
In Lemma \ref{l1} we show that  there exist constants $\clk$ such that 
\eq{rnmm} \ji \|\kj b |x|^m \Psi \|_\hhh^2 \dk\leq \YYY, 
\en 
Combining  \kak{R2} and \kak{rnmm},  
we see in Lemma \ref{l2}
that 
$$\ji  \|\kj a \Psi\|_\hhh^2 \dk$$
\eq{ke}
\leq 
2^n \lkk\ji \|\kj b \Psi \|_\hhh^2 \dk+
\ZZZ \R_{n-l,l}(\Psi)\rkk 
\en 
with some constants $\dln$, 
where 
$$\R_{n,m}(\Psi)=
\YYY.$$
Furthermore    {\it if } $\gr\in D(\N)$ then 
we  see that 
$$\N\gr=\eht \ee\N\gr+\ee[\N,\eht]\gr,$$
where 
$$ 
E=\is (H).
$$ 
Using this identity 
we show in Lemma \ref{ess2} that  
if $\gr\in D(\N)$ then 
for all $l\geq 0$, 
\eq{ttl}
|x|^l\gr\in D(\N).
\en
Under these preparations 
we prove Theorem \ref{t1} by means of an induction. 
Let us assume that 
\eq{ind}
\gr \in D(N^{(n-1)/2}).
\en 
Hence 
\eq{TOG}
\ji \|\kl a\gr\|^2_\hhh\dkl<\infty,\ \ \ l=1,2,...,n-1.
\en 
Then we see that by \kak{ind}, 
$$
\ZZZ    \R_{n-l,l}(\gr)<\infty.$$
Moreover by using  pull through formula \kak{23} 
we prove in Lemma \ref{sin1} that 
$$\|\kj b\gr\|_\hhh\leq 
\sump \delta_1(k_p) \|b(\k_1)...
\widehat{b(\k_p)}...b(\k_n)(|x|+1) \gr\|_\hhh$$
\eq{100}
+
\sump \sum_{q<p}
\delta_2(k_p, k_q) \|b(\k_1)...\widehat{b(\k_q)}...\widehat{b(\k_p)}...b(\k_n)|x|^2\gr\|_\hhh
\en 
with 
$$\delta_1\in\LR,\ \ \ \delta_2\in L^2(\BR\times \BR).$$
By \kak{rnmm},  \kak{ttl} and  assumption \kak{ind}, 
 we show that 
$$ 
\ji \|\kj b\gr\|_\hhh \dk<\infty.
$$
Hence by 
\kak{ke} 
we have 
$$\ji  \|\kj a \gr\|_\hhh^2 \dk$$
$$ 
\leq 
2^n \lkk\ji \|\kj b \gr \|_\hhh^2 \dk+
\ZZZ   \R_{n-l,l}(\gr)\rkk<\infty, 
$$ 
which implies, together with \kak{TOG}, that 
 $$\gr\in  D(N^{n/2}).$$ 
Since 
 $\gr\in D(N^\han)$ is known, 
we obtain 
$$\gr\in \bigcap_{k=1}^\infty D(\N).$$

This paper is organized as follows. In Section 2 
we establish \kak{100} by means of the pull-through formula. 
In Section 3 we 
give a proof of the main theorems. 
In Section 4 we show  \kak{ttl} by virtue of 
a functional integral representation.

\section{Pull-through formula and exponential decay} 
\subsection{Fundamental facts}
Let $T$ be an operator. 
We set 
$$C^\infty(T)=\bigcap_{k=1}^\infty D(T^k).$$ 
\bl{p2}
We have  $\gr\in  C^\infty(|x|)\cap D(\Delta) \cap C^\infty(\hf)$.
\el
\proof
By Proposition \ref{hiroshima}, $D(H)=D(\Delta)\cap D(\hf)$.
Then $\gr\in D(H)$, which implies 
$\gr\in D(\Delta)$. 
By Proposition \ref{p1} (2) it holds that $\gr\in C^\infty(|x|)$. 
It is obtained in \cite{fgs} that 
$\hf^l (H-i)^{-l}$ is bounded for all $l\geq 0$. 
Recall that $E=\is(H)$. 
Then 
it follows that for arbitrary $l\geq 0$,  
$$\|\hf^l\gr\|_\hhh 
=\|\hf^l (H-i)^{-l}(E-i)^l\gr\|_\hhh
\leq 
|(E-i)^l| 
\|\hf^l (H-i)^{-l}\|
\|\gr\|_\hhh.$$
Then $ \gr\in D(\hf^l)$ for all $l\geq 0$. 
Thus the lemma follows.
\qed
Let 
$$\fffw={\cal L} \{a^\ast(f_1,j_1)...a^\ast(f_n,j_n)\Omega, \Omega| f_j\in C_0^\infty(\BR), 
j=1,...,n, n=0,1,...\},$$
where ${\cal L}\{...\}$ denotes  the set of the finite linear sum of $\{...\}$. 
We define 
$$\d=C^\infty(|x|) \cap  C^\infty(\hf),$$ 
and 
$$\EE=\ccc\widehat{\otimes} \fffw.$$

\bl{essen}
Let  $m\geq 0$ and $n\geq 0$. Then 
$\hff^n+|x|^m$ is self-adjoint on $D(\hff^n)\cap D(|x|^m)$ 
and essentially self-adjoint on $\EE$. 
\el
\proof 
The self-adjointness is trivial. 
Since $\ccc$ and $\fffw$ are the set of analytic vectors of $|x|^m$ and 
$\hff^n$ respectively, 
$\ccc$ and $\fffw$ are    cores of $|x|^m$ and 
$\hff^n$ respectively. 
Hence $\EE=\ccc\widehat{\otimes}\fffw$ is a core of 
$\hff^n+|x|^m$.  
\qed

\begin{remark}
Let $p,q\geq 0$. From Lemma \ref{essen} it follows that for 
$\Psi\in \d\subset D(\hff^p+|x|^q)$ there exists a sequence $\Psim\subset \EE$ such that 
$\Psi_m\rightarrow\Psi$ and $(\hff^p+|x|^q)\Psi_m\rightarrow (\hff^p+|x|^q)\Psi$ strongly as $m\rightarrow \infty$. 
\end{remark}
Let $f_j\in\ccccc$, $j=1,...,n$, and $\Psi\in \EE$. 
Then it is  well known and easily proven  that 
\eq{o}
 \ji  |\F| \| \kj a \Psi\|_\hhh \dk 
 \leq \e(f_1,...,f_n) \|(\hf+1)^{n/2}\Psi\|_\hhh
\en 
with some constant $\e(f_1,...,f_n)$ independent of $\Psi$. 

Let $A$ and $B$ be operators. 
We say $f\in D(AB)$ if $f\in D(B)$ and $Bf\in D(A)$.

\bl{xa}
Let $\Psi\in \d$. 
Then there exists  ${\cal M}_\d(\Psi)\subset\RR^{3n}$ with the Lebesgue measure zero  
such that 
\eq{KK}
\Psi\in D(\kj a)
\en and 
\eq{KKK}
\kj a \Psi\in\d
\en 
for $(\kn)\not\in {\cal M}_\d(\Psi)$. 
Moreover assume that $\Psim\subset\EE$ satisfies that $\Psi_m\rightarrow \Psi$ and $\hff^{n/2}\Psi_m\rightarrow \hff^{n/2}\Psi$ strongly as 
$m\rightarrow \infty$. 
Then there exists a subsequence $\mdd\subset \md$ and 
${\cal M}_{\d}(\Psi, \Psim,\mdd)\subset\RR^{3n}$ with the Lebesgue measure 
zero such that 
for $(\kn)\not\in {\cal M}_{\d}(\Psi,\Psim, \mdd)$, 
\kak{KK} and \kak{KKK} are valid and 
$$s-\!\!\limd \kj a \Psid=\kj a \Psi.$$
\el
\proof 
See Appendix A. 

\bl{x}
The operator $|x|$ leaves $\d$ invariant. 
\el
\proof
Let $\Psi\in \d$. 
It is clear that $|x|\Psi\in \cx$. 
We choose a sequence $\{\Psi_m\}\subset \EE$ 
such that $\Psi_m\rightarrow \Psi$ and 
$(\hff^{2n}+|x|^2)\Psi_m\rightarrow 
(\hff^{2n}+|x|^2)\Psi$ strongly as $m\rightarrow \infty$. 
In particular 
\eq{pa}
|x|\Psi_m\rightarrow |x|\Psi
\en  strongly 
as $m\rightarrow \infty$. 
$\hf^n |x|\Psi_m$ is well defined and it is obtained that 
$$\|\hf^n |x|\Psi_m\|^2_\hhh\leq \|\hf^{2n}\Psi_m\|_\hhh
\||x|^2\Psi_m\|_\hhh
\leq \|(\hff^{2n}+|x|^2)\Psi_m\|_\hhh^2.$$
Then $\hf ^n |x|\Psi_m$ converges strongly as $m\rightarrow \infty$. 
Since  $\hf^n$ is closed, by \kak{pa} 
we have  $|x|\Psi\in D(\hf^n)$. 
Here $n$ is arbitrary, hence $|x|\Psi\in \ch$. 
The proof is complete. 
\qed

Let 
$$\beta(\k)=\frac{e}{\sqrt 2}\frac{\kxx}{\sqrt{\omega(k)}}e(\k)\vp(k)$$
and 
$$b(\k)=\eaa a(\k) \ea  = a(\k)-ix\cdot \beta(\k).$$
For simplicity we set $-ix\cdot \beta(\k_j)=\t_j$. 
Then 
$$b(\k_j)=a(\k_j)+\t_j.$$

\bl{xxa}
Let $\Psi\in\EE$ and 
$f_j\in\ccccc$, $j=1,...,n$. Then 
there exists a constant 
$\e'(f_1,...,f_n)$ independent of $\Psi$ such that 
\eq{op}
\ji  |\F| \| {\kj b} \Psi\|_\hhh \dk 
\leq \e'(f_1,...,f_n)
\| \KKK n  \Psi\|_\hhh.
\en 
\el
\proof 
Since $[\theta_j, a(\k)]=0$ on $\EE$,  we have 
 $$\kj b \Psi= (a(\k_1)+\t_1) ...  (a(\k_n)+\t_n) \Psi$$
$$ =\xxx \tl \kjj a  \Psi .$$
Hence by \kak{o},   
$$\ji  |\F|\| \kj b \Psi \|_\hhh \dk $$
$$ \leq \xxx   \tll$$
$$\times  
\e(f_1,..,\widehat{f_{p_1}},..,\widehat{f_{p_l}},...,f_n) \| (\hf+1)^{(n-l)/2} |x|^l\Psi\|_\hhh.$$
Since 
$\| (\hf+1)^{(n-l)/2} |x|^l\Psi\|_\hhh\leq c_{nl}
\|\KKK n\Psi\|_\hhh$ 
with some constant $c_{nl}$. 
Thus \kak{op}  follows. 
\qed
\bl{ae}
Let $\Psi\in \d$. 
Then there exists  ${\cal N}_{\d}(\Psi)\subset\RRR$ 
with the Lebesgue measure zero such that 
\eq{OP}
\Psi\in D(\kj b)
\en  and 
\eq{OPP}
\kj b \Psi\in \d
\en 
for $(\kn)\not\in{\cal N}_{\d}(\Psi)$. 
Moreover assume that $\Psim\subset\EE$ satisfies that 
$\Psi_m\rightarrow \Psi$ and 
$(\hff^n +|x|^{2n})\Psi_m\rightarrow (\hff^n +|x|^{2n}) 
\Psi$ strongly as 
$m\rightarrow \infty$. 
Then there exists a subsequence $\mdd\subset \md$ and 
${\cal N}_{\d}(\Psi, \Psim,\mdd)\subset\RR^{3n}$ 
with the Lebesgue measure zero such that 
for $(\kn)\not\in {\cal N}_{\d}(\Psi,\Psim, \mdd)$, 
\kak{OP} and \kak{OPP} are valid and 
$$s-\!\!\limd \kj b \Psid=\kj b \Psi.$$
\el
\proof 
See Appendix A.

\subsection{Pull-through formula}

\bl{tt}
We have 
\eq{ch}
\EE  \subset  D(H\kj b) \cap D(\kj b H)
\en  for all $\kr$, 
and for $\Psi\in\EE$, 
$$[H, \kj b]\Psi
=\S 0 \Psi+\S 1  \Psi+\S 2  \Psi.$$
Here 
$$\S 0={\cal R}_0(\k_1,...,\k_n)= 
-\sump \omega(k_p) \kj b,  $$
$$\S 1 ={\cal R}_1(\k_1,...,\k_n)=  
\sump \HH (\k_p) 
b(\k_1) ...  \widehat{b(\k_p)} ...  b(\k_n),$$
$$\S 2 ={\cal R}_2(\k_1,...,\k_n)=   
\sump \sum_{q<p} \HHH(\k_p,\k_q) 
b(\k_1) ...  \widehat{b(\k_q)} ...  
\widehat{b(\k_p)} ...  b(\k_n), 
$$
and 
$$\HH(\k)=\HH(\k,x,p)=\frac{i}{2}\lkk 
(x\cdot\beta(\k)) k \cdott(p-eA)+k\cdott (p-eA)(x\cdot\beta(\k))\rkk-i\omega(k)(x\cdot\beta(\k)),$$
$$\HHH(\k,\k')=\HHH(\k,\k',x)=
(x\cdot \beta(\k)) (x\cdot \beta(\k'))(k\cdot k').$$
\el
\proof
\kak{ch}  is trivial. 
On $\EE$ we have 
\eq{21}
[H, b(\k)]=-\omega(k) b(\k) +\HH(\k).
\en 
Moreover 
\eq{26}
[b(\k'), \HH(\k)]= \HHH(\k,\k').
\en 
By \kak{21} and   \kak{26} we have 
$$[H, \kj b] \Psi 
=
\sump b(\k_1) ...  
\lkk -\omega(k_p)b(\k_p)+\HH(\k_p) \rkk  ...  b(\k_n) \Psi$$
$$=
-\sump \omega(k_p) \kj b  \Psi
+
\sump \HH(k_p) 
b(\k_1) ...  \widehat{b(\k_p)} ...  b(\k_n)  \Psi$$
$$+\sump \sum_{q<p} \HHH(\k_p,\k_q) 
b(\k_1) ...  \widehat{b(\k_q)} ...  
\widehat{b(\k_p)} ...  b(\k_n)  \Psi.$$
The lemma follows.
\qed

$\ov B$ denotes the closure of $B$. We simply set $\ov{\S 1 }=\ov{\S 1 \lceil_{\EE}}$. 
\bl{nn1}
Let $\Psi\in \d\cap D(\Delta)$. Then there exists 
$\nnd\subset \RR^{3n}$ with the Lebesgue measure zero such that for $(\kn)\not\in\nnd$, 
$$\Psi\in  D(\S 0 (\knn))\cap D(\ov{\S 1 }(\knn))
\cap D(\S 2 (\knn)).$$ 
\el
\proof 
By Lemma \ref{ae}, $\Psi\in D(\kj b)$ and 
$ \kj b\Psi\in\d$ for $(\kn)\not\in\nd$. 
Thus $\kj b\Psi\in D(\sump \omega(k_p))\cap D(\HHH(\k_p,\k_q))$, 
which implies that 
$\Psi\in   D(\S 0(\knn))\cap D(\S 2(\knn) )$. 
Next we shall prove 
$D(\ov{\S 1 }(\knn))\ni \Psi$. 
Simply we set $\KK_n=\KKK n$. 
We have on $\EE$ 
$$\S 1 = \sump ix\cdot\beta(\k_p)(k_p\cdot p) \kkj b+ \sump \rx \kkj b,$$
where $$\rx=
(-ie)(x\cdot\beta(\k_p)) (k\cdot A)
-
\frac{i}{2}(i\beta(\k_p)\cdot k_p+x\cdot\beta(\k_p)|k_p|^2)
-i\omega(k_p)(x\cdot\beta(\k_p)). $$ 
It follows that for $\Phi\in\EE$,  
$$ 
\ji  \|\F  \rx \kkj b \Phi \|_\hhh \dk \leq \cccc 1$$
with some constants $c_1$ and $m_1$, and 
$$ix\cdot\beta(\k_p) (k_p\cdot p) \kkj b\Phi =
ix\cdot\beta (\k_p)  \kkj b (k_p\cdot p)  \Phi$$
\eq{se}
+
\sum_{q\not =p} \rrx 
\kkkj b  \Phi,
\en 
where 
$$\rrx=ix\cdot\beta(\k_p) \lk 
-k_q\beta(\k_q)+ix\cdot \beta(\k_q)(k_p\cdot k_q) \rk.$$ 
The second term of \kak{se}  is estimated as 
$$
\ji  \|\F  \sum_{q\not =p} \rrx 
\kkkj b  \Phi\|_\hhh\dk 
$$
$$\leq \cccc 2$$ 
with some constants $c_2$ and $m_2$. 
By \kak{op} the first term of \kak{se} is estimated as 
$$	
\ji  \|\F \lk x\cdot\beta(\k_p)\rk  \kkj b (k_p\cdot p) \Phi\|_\hhh\dk $$
$$\leq 
\epsilon'(f_1,...,\widehat{f_p},...,f_n)
\int  |f(k_p)| \frac{|e|}{\sqrt 2}\frac{|\vp(k_p)|}{\sqrt{\omega(k_p)}}  
\|\KK_{n-1}|x|  (k_p\cdot p) \Phi\|_\hhh dk_p. $$ 
Let $Q=\KK_{n-1}$. 
 Note that 
$$\|Q  |x|  (k\cdot p) \Phi\|_\hhh^2=
(|x|^2 Q^2 \Phi, (k_p\cdot p)^2\Phi)_\hhh
+
(\Phi, [(k_p\cdot p), Q^2|x|^2](k_p\cdot p)\Phi)_\hhh.$$
Since $[(k_p\cdot p), |x|]=-i (k_p\cdot x)/|x|$, 
we have 
$[(k_p\cdot p), Q^2|x|^2]=k_p\cdot P,$
where 
$$P =2\lkk (\hf+1)^{n-1}(-i)\frac{x}{|x|}+(-i) x (|x|+1)^{2n-3}+(-i)\frac{x}{|x|}(|x|+1)^{2n-2}\rkk.$$ 
Then 
$$\|Q |x| (k_p\cdot p) \Phi\|_\hhh^2
\leq 
|k_p|^2\lk 
\||x|^2 Q ^2 \Phi\|_\hhh\|\Delta \Phi\|_\hhh+ 
\|P \Phi\|_\hhh \||p|\Phi\|_\hhh\rk.$$
Hence 
$$\|Q |x| (k_p\cdot p) \Phi\|_\hhh\leq 
|k_p| \lk  \cccc 3+c'\|\Delta \Phi\|_\hhh \rk $$ 
follows with some constants $c_3$, $c'$  and $m_3$. 
Thus 
for $\Phi\in\EE$ 
\eq{rj}
\ji   \|\F \S{1}(\knn)\Phi\|_\hhh \dk  \leq 
 c  \|\KK_m \Phi\|_\hhh+c'\|\Delta \Phi\|_\hhh 
\en 
follows with some constants $c$ and $m$. 
Set  $K=-\Delta+\KK_m= -\Delta+|x|^{2m}+\hff^m$. 
Then $K$ is self-adjoint on $D(-\Delta+|x|^{2m})\cap D(\hff^m)$ and essentially self-adjoint on 
$\EE$. Then for $\Psi\in\d\cap D(\Delta)$ there exists a sequence 
$\{\Psi_l\}\subset  \EE$ such that 
$\Psi_l\rightarrow \Psi$ and $K\Psi_l\rightarrow K\Psi$ strongly 
as $l\rightarrow \infty$. 
By \kak{rj}  it follows that 
$$
\ji   \|\F \S{1}(\knn)\Psi_l\|_\hhh \dk  
\leq 
 c  \|\KK_m \Psi_l\|_\hhh+c'\|\Delta \Psi_l\|_\hhh.$$
Then there exist $\nd'\subset\RR^{3n}$ with the Lebesgue measure zero 
and 
 a subsequence $\{l'\}\subset\{l\}$ such that 
$
\S{1}(\knn )\Psi_{l'}$ strongly converges as $l'\rightarrow \infty$ 
for $(\kn)\not\in\nd'$. 
Then 
$\Psi\in D(\ov{\S 1 }(\knn))$ for $(\kn)\not\in\nd'$. 
Set $$\nnd=\nd\cup\nd'.$$
We get the desired results. 
\qed

The following lemma is a variant of the pull-through formula. 
\bl{31}
For $(\kn)\not \in\g$, the following (1), (2) and (3) hold; 
\bi
\item[(1)] 
$\gr\in D(\kj b )\cap D(\S 0)\cap D(\ov{\S 1 })\cap D(\S 2 ),$  
\item[(2)] $\kj b  \gr\in D(H )$, 
\item[(3)] 
\eq{oi}
\lk H-E+\sw \rk  \kj b  \gr 
=\ov{\S 1 } \gr+{\S 2 } \gr.
\en 
\ei
In particular it follows that for  $(\kn)\not \in\g$ and $(\kn)\not=(0,...,0)$, 
\eq{23}
\kj b \gr=
\lk H-E +\sump \omega(k_p)\rk \f \lk \ov{\S 1 }\gr+{\S 2 }\gr\rk.
\en 
\el
\proof 
Note that $\gr\in \d \cap D(\Delta) = 
C^\infty(|x|) \cap C^\infty(\hf)\cap D(\Delta)$. 
Then (1) follows from Lemma  \ref{nn1}.
Since $\EE$ is a core of $H$, we have $\phi_m\in\d$ such that 
$\phi_m\rightarrow\gr$ and $H\phi_m\rightarrow H\gr=E\gr$ strongly as $m\rightarrow\infty$. 
Then we have for $\phi\in\EE$ 
$$(H\phi, \kj b \phi_m)_\hhh =\jii (\phi, {\cal R}_j\phi_m)_\hhh +(\phi, \kj b H \phi_m)_\hhh .$$
It follows that 
$$\limm (H\phi, \kj b \phi_m)_\hhh =\limn (\jk b H\phi, \phi_m)_\hhh $$
$$=(\jk b H\phi, \gr)_\hhh =(H\phi, \kj b \gr)_\hhh ,$$
$$\limm (\phi, {\cal R}_j\phi_m)_\hhh =
\limm ({\cal R}_j^\ast \phi, \phi_m)_\hhh =({\cal R}_j^\ast \phi, \gr)_\hhh =(\phi, \ov{{\cal R}_j}\gr)_\hhh ,$$
and 
$$\limm (\phi, \kj b H \phi_m)_\hhh =\limn (\jk b\phi, H\phi_m)_\hhh $$
$$=(\jk b\phi, E\gr)_\hhh =
E (\phi, \kj b \gr)_\hhh .$$
Hence 
$$(H\phi, \kj b \gr)_\hhh =\jii (\phi, \ov{{\cal R}_j}\gr)_\hhh +E(\phi, \kj b \gr)_\hhh .$$
Then 
$\kj b \gr\in D(H)$ and we have 
$$H\kj b \gr=\jii \ov{{\cal R}_j} \gr+E \kj b \gr.$$
Note that $\ov{{\cal R}_0}\gr={\cal R}_0\gr $ and $\ov{{\cal R}_2}\gr={\cal R}_2\gr$. 
Then \kak{oi} follows. 
\qed

\subsection{Exponential decay of $\N\gr$} 
\bl{ess}
Suppose that $\gr\in D(\N)$. Then there exist positive constants $D_k$, 
 and $\delta$ independent of $k$  
such that 
$$\|N^{k/2}\gr(x)\|_\fff \leq D_k e^{-\delta|x|^{m+1}}$$
for almost every $x\in\BR$. 
In particular 
$\N \gr\in D(e^{\delta |x|^{m+1}})$. 
\el
The proof of Lemma \ref{ess} is based on a functional integral representation of 
$e^{-tH}$. 
Essential ingredients of the   proof  have been obtained  in 
\cite{h11}. 
The proof is, however, long and complicated. 
Then we move it  to Appendix B.    
\bl{ess2}
Suppose that $\gr\in D(\N)$. Then $|x|^l\gr\in D(\N)$ for all $l\geq 0$. 
\el
\proof 
This lemma is immediately follows from 
Lemma \ref{ess} and  the following fundamental lemma. 
\qed
\bl{f2}
Let ${\cal K}$  be  a Hilbert space,  and 
$A$ and $B$ self-adjoint operators such that 
$[e^{-its A}, e^{-is B}]=0$ for $s,t\in\RR$. 
Suppose that $\phi\in D(A)\cap D( B)$ and 
$ A\phi\in D(B)$. Then
$ B\phi\in D(A)$ with 
$ 
AB \phi 
=BA\phi$. 
\el
\proof 
It follows that 
$t\f (e^{-it A}-1)e^{-is B}\phi=
t\f e^{-is B} ({e^{-it A}-1}) \phi$.  
Take $t\rightarrow 0$ on the both sides. Then 
it follows that $e^{-is B}\phi\in D(A)$ with 
$
A e^{-is B}\phi= e^{-is B}  A\phi.$
From this identity  we have 
$s\f A ({e^{-is B}-1}) 
\phi= s\f ({e^{-is B}-1})   A \phi.$
Take $s\rightarrow 0$ on the both sides. Since $A$ is closed and 
assumption $ A\phi\in D(B)$, 
we see that 
$ B \phi\in D(A)$  and $AB\phi=BA\phi$. 
\qed

{\it Proof of Lemma \ref{ess2}}

In Lemma \ref{f2} we set ${\cal K}=\hhh$, $A=\N$ and $B=|x|^l$. 
Since $\gr\in  D(\N)\cap D(|x|^l)$ and $\N \gr\in D(|x|^l)$ by Lemma \ref{ess}, the lemma follows. 
\qed

\section{Proof of  the main theorems}
\bl{b1}
The following statements are equivalent. 
\bi
\item[(1)]  $\Psi\in 
D(\kj a)$ for almost every $\kr$ and  
\eq{27} \ji \|\kj a \Psi \|^2_\hhh\dk<~\infty.
\en
\item[(2)]
$\Psi\in D (\prod_{j=1}^n (N-j+1)^\han)$.
\ei
Moreover if (1) or (2) is satisfied, then 
it holds that 
$$\ji \|\kj a \Psi \|_\hhh^2 \dk=\|\prod_{j=1}^n (N-j+1)^\han\Psi\|_\hhh^2.$$
\el
\proof
We prove $(1)\Longrightarrow (2)$. 
We identify $\hhh$ as 
\eq{iden}
\hhh\cong \bigoplus_{n=0}^\infty \hhnn,
\en 
where 
$$\hhnn= L^2(\BR\times (\BR\times\{1,2\})_{\rm sym}^{\times n}).$$
 We note that 
$$\lk \prod_{j=1}^n (N-j+1)^\han\Psi\rk^{(k)}=\lkk
\begin{array}{ll}
0,& k=0,1,...,n-1,\\
\sqrt{k(k-1)...(k-n+1)}\Psi^{(k)},&k\geq n.
\end{array}
\right.
$$
Define $\Psi_p=\oplus_{m=0}^\infty \Psi_p^{(m)}\in\hhh$ 
by 
\eq{P}
\Psi_p^{(m)}=\lkk\begin{array}{ll}\Psi^{(m)},& m\leq p,\\
0,&m>p.\end{array}\right.
\en 
By the definition of $a(\k)$ we have 
$$\lk \kj a \Psi_p\rk^{(l)}(x, \m_1,...,\m_l)=
\sqrt{l+1}\sqrt{l+2}...\sqrt{l+n}\Psi_p^{(l+n)}(x, \k_1,...,\k_n, \m_1,...,\m_l).$$
Then 
$$\|  \kj a \Psi_p \|_\hhh^2 $$
$$=
\sum_{l=0}^\infty (l+1)(l+2)...(l+n) \ji \| \Psi_p^{(l+n)}
(\cdot, \k_1,...,\k_n,\m_1,...,\m_l)\|_\LR^2 \dm$$
$$ 
=
\sum_{l=0}^{p-n}  (l+1)(l+2)...(l+n) \ji \| \Psi^{(l+n)}
(\cdot, \k_1,...,\k_n,\m_1,...,\m_l)\|_\LR^2 \dm
$$
\eq{K1}
=\sum_{l=0}^{p-n}\|(\kj a\Psi)^{(l)}\|^2_{\hhh_l}.\en 
Hence 
$$\ji \|\kj a\Psi_p\|^2_\hhh \dk$$
$$=\sum_{l=0}^{p-n}  (l+1)(l+2)...(l+n) 
\ji  \|\Psi^{(l+n)}(\cdot, \k_1,...,\k_n,\m_1,...,\m_l)\|_\LR^2 \dm\dk$$
$$=\sum_{k=n}^p k(k-1)(k-2)...(k-n+1)
\ji  \|\Psi^{(k)}(\cdot, \k_1,...,\k_k)\|_\LR^2 dk_1...dk_k$$
\eq{K2}
 =\sum_{k=0}^p 
\left \|\lk \prod_{j=1}^n (N-j+1)^\han\Psi\rk^{(k)}\right\|^2_{\hhh_k}. 
\en 
By  (1) we see that 
$$
\limp \|\kj a\Psi_p\|_\hhh^2=\limp\sum_{k=0}^{p-n} \|(\kj a\Psi)^{(k)}\|^2_{\hhh_k}$$
\eq{K3}
=\|\kj a \Psi\|^2_\hhh
\en 
for almost every $(\kn)\in \RR^{3n}$, and  
$$\|\kj a \Psi\|^2_\hhh\in L^1(\RR^{3n}).$$
Thus the Lebesgue dominated convergence theorem yields that 
$$\limp \ji\|\kj a\Psi_p\|^2_\hhh \dk <\infty.$$
Then 
from  \kak{K2}, it follows that 
$$ 
\left \| \prod_{j=1}^n (N-j+1)^\han\Psi \right\|^2_\hhh=
\limp 
\sum_{k=0}^p 
\left \|\lk \prod_{j=1}^n (N-j+1)^\han\Psi\rk^{(k)}\right\|^2_{\hhh_k} <\infty.$$
Thus (2)  follows.

We prove $(2)\Longrightarrow (1)$. 
By \kak{K2} and  (2) we see that 
$$\limp \ji \|\kj a\Psi_p\|^2_\hhh \dk<\infty, $$
and by \kak{K1},  $\|\kj a\Psi_p\|^2_\hhh$ is increasing in $p$. 
Then we have by the Lebesgue monotone  convergence theorem, 
\eq{dom}
\limp  \ji  \|\kj a\Psi_p\|^2_\hhh \dk
=
 \ji \limp \|\kj a\Psi_p\|^2_\hhh \dk<\infty .
\en 
Then (1) follows from 
\kak{K3}. 
\qed

\bl{b3}
The following statements are equivalent. 
\bi
\item[(1)]  $\Psi\in  D(\kj a)$ 
for almost every $\kr$ and
$$ \ji \|\kj a \Psi \|^2_\hhh\dk<~\infty$$ 
for $n=1,2,...,k. $ 
\item[(2)]
$\Psi\in D (N^{k/2})$. 
\ei
\el
\proof
By Lemma \ref{b1}, it is enough to show that 
$$
 \bigcap_{k=1}^n D(\prod_{j=1}^k (N-j+1)^\han )=D(\N). 
$$ 
Assume that 
\eq{22} \Psi\in 
 \bigcap_{k=1}^n D(\prod_{j=1}^k (N^n-j+1)^\han ).
\en 
Let $\Psi_p$ be defined by \kak{P}. 
Let $W_n=\prod_{j=1}^n (N-j+1)$.  For example 
$N=W_1, N^2=W_2+W_1, N^3=W_3+3W_2+W_1, N^4=W_4+6W_3+7W_2+W_1,$ etc.
One can inductively   see that there exist constants $a_j, j=1,...,k$,  
such that on $\fff_0$, 
$$N^k=\sum_{j=1}^n a_j W_j.$$ 
Then it follows that 
\eq{nnn} 
\|N^{k/2}\Psi_p\|_\hhh^2 =a_1 \|  W_1^\han\Psi_p\|_\hhh^2   + a_2 
\| W_2^\han   \Psi_p\|_\hhh^2 +
 ...  +a_k\|  W_k^\han \Psi_p\|_\hhh^2   .
\en 
As $n\rightarrow \infty$,  from \kak{22} it follows that 
the right hand side of \kak{nnn} converges to 
$$
a_1 \|  W_1^\han\Psi\|_\hhh^2   + a_2 
\| W_2^\han  \Psi\|_\hhh^2 +
 ...  +a_k\|  W_k^\han\Psi\|_\hhh^2 .
$$
Since 
$$\|\N\Psi\|_\hhh^2=\limp \sum_{k=0}^p \|(\N\Psi)^{(k)}\|_{\hhh_k}^2=\limp\|\N\Psi_p\|_\hhh^2<\infty,$$
$\Psi\in D(N^{k/2} )$ follows. 
Then 
\eq{OO}
 \bigcap_{k=1}^n D(\prod_{j=1}^k (N^n-j+1)^\han )\subset D(\N).
\en 
Next we assume that 
$$\Psi\in D(\N).$$
Note that 
$$\Psi\in \bigcap_{l=1}^k D(N^{l/2}).$$
It is seen that there  exist   constants $\bln$, $l=1,...,n$,  such that 
$$ 
\|\prod_{j=1}^n(N-j+1)^\han\Psi_p\|^2 
=(\Psi_p, \prod_{j=1}^n(N-j+1)\Psi_p)
$$
\eq{bound}
=(\Psi_p, N(N-1)(N-2)\cdots (N-n+1)\Psi_p)\leq 
\sum_{l=1}^n \bln \|N^{l/2}\Psi_p\|^2.
\en 	
Take $p\rightarrow \infty$ on the both sides above. 
Then the right hand side of \kak{bound} converges to 
$$\sum_{l=1}^n \bln \|N^{l/2}\Psi\|^2.$$
Hence 
$$\|\prod_{j=1}^n(N-j+1)^\han\Psi\|_{\hhh}^2=
\limp \sum_{k=0}^p \|\lk \prod_{j=1}^n(N-j+1)^\han\Psi\rk^{(k)} \|_{\hhh_k}^2$$
$$
=
\limp \|\prod_{j=1}^n(N-j+1)^\han\Psi_p\|_\hhh^2<\infty.$$
Thus $\Psi\in \bigcap_{k=1}^n D(\prod_{j=1}^n(N-j+1)^\han)$. 
We obtain 
\eq{OOO}
\bigcap_{k=1}^n D(\prod_{j=1}^n(N-j+1)^\han)\supset D(\N).
\en 
The lemma follows from \kak{OO} and \kak{OOO}.
\qed

We set 
$$\HP=\HP(\kn)=\lk H-E+\sump \omega(k_p)\rk\f.$$

\bl{dia}
There exist $\delta_1(\cdot)\in\LR$ and $\delta_2(\cdot,\cdot)\in L^2(\BR\times\BR)$ 
such that for  $\Psi\in\d$, 
\eq{200}
\|\ov{\HP \HH(\k_q)}\Psi\|_\hhh\leq \delta_1(k_q)\|(|x|+1)\Psi\|_\hhh,
\en 
and 
\eq{201}
\|\ov{\HP \HHH(\k_q,\k_p)}\Psi\|_\hhh\leq \delta_2(k_q,k_p)\||x|^2\Psi\|_\hhh.
\en 
\el
\proof 
By the closed graph theorem there exists a constant $C$ such that 
$$\|(-\Delta + \hf)\Psi\|_\hhh\leq C\|(H+1)\Psi\|_\hhh.$$
First we shall prove that 
$\ov{\HP (p\cdot k_q)}$ and 
$\ov{\HP (A\cdot k_q)}$ are bounded   with 
\eq{000}
\|\ov{\HP (p\cdot k_q)}\|\leq \c(k_q)
\en 
 and 
\eq{0000} \|\ov{\HP (A\cdot k_q)}\|\leq \cc(k_q),
\en 
where 
$\c(k_q)=\sqrt{(|k_q|+|1+E|)C}$ 
and
$ \cc(k_q)={\sqrt2} \lk \c(k_q) +1\rk (2\|\vp/\omega\|+\|\vp/\sqrt\omega\|).$
Let $\Psi\in\EE$. 
Since 
$$\|(p\cdot k_q)\Psi\|_\hhh^2\leq |k_q|^2(\Psi, C(H+1)\Psi)\leq |k_q|^2 C\lkk 
\|(H-E)^\han \Psi\|_\hhh^2 +|1+E|\|\Psi\|_\hhh^2\rkk ,$$
we see that 
$$\|(p\cdot k_q) \HP \Psi\|_\hhh^2\leq C|k_q|\|\Psi\|_\hhh^2+C|1+E|\|\Psi\|_\hhh^2.$$
Thus \kak{000} follows.
Note that 
$$\|a(f,j)\Psi\|_\hhh
\leq \|f/\sqrt\omega\| \|\hf^\han\Psi\|_\fff,$$
and 
$$
\|\add (f,j)\Psi\|_\fff 
\leq \|f/\sqrt\omega\| \|\hf^\han\Psi\|_\fff+\|f\|\|\Psi\|_\fff.$$
Since 
$$\|(A\cdot k_q)\Psi\|_\hhh \leq 
\sqrt 2
|k_q| 
\lk 2\|\vp/\omega\|+\|\vp/\sqrt\omega\|\rk  \lk \|\hf^\han\Psi\|_\hhh+\|\Psi\|_\hhh\rk$$
and 
$$\|\hf^\han\Psi\|_\hhh^2\leq C(\Psi, (H+1)\Psi)_\hhh\leq C\|(H-E)^\han\Psi\|_\hhh^2 +C|1+E|\|\Psi\|_\hhh^2,$$
we have 
$$|k_q|^2\|\hf^\han\HP\Psi\|_\hhh^2\leq 
C|k_q|\|\Psi\|_\hhh^2+C|1+E|\|\Psi\|_\hhh^2.$$
Hence 
$$\|(A\cdot k_q) \HP \Psi\|_\hhh\leq 
\sqrt 2 \lk 2\|\vp/\omega\|+\|\vp/\sqrt\omega\|\rk 
 \lk 
|k_q| \|\hf^\han\HP\Psi\|_\hhh+|k_q|\|\HP\Psi\|_\hhh\rk$$
$$\leq \sqrt 2 \lkk \sqrt{(|k_q|+|1+E|)C}+1\rkk (2\|\vp/\omega\|+\|\vp/\sqrt\omega\|)\|\Psi\|_\hhh.$$
Thus \kak{0000}  follows. 
We have on $\EE$
$$\HH(\k)=i(p-eA)\cdot k(  x\cdot\beta(k))  +
\frac{i}{2}\lk i\beta(\k)\cdot k+x\cdot\beta(\k)|k|^2\rk-i\omega(k)(x\cdot\beta(\k)).$$
Then by \kak{000} and \kak{0000} 
we have for $\Psi\in\EE$,  
$$\|\HP i(p-eA)\cdot k_p, (x\cdot\beta(k_p))\Psi\|_\hhh
\leq 
\lk \c(k_p)+|e|\cc(k_p)\rk \te \frac{|\vp(k_p)|}{\omega(k_p)} \||x|\Psi\|_\hhh,$$
$$\|\HP (-i\omega(k_p)(x\cdot\beta(\k_p))) \Psi\|_\hhh
\leq \te \sqrt{\omega(k_p)}|\vp(k_p)|\||x|\Psi\|_\hhh,$$
and 
$$\|\HP \frac{i}{2}\lk i\beta(\k_p)\cdot k_p+x\cdot\beta(\k_p)|k_p|^2\rk\Psi\|_\hhh
\leq \half\te\lk 
\frac{|\vp(k_p)|}{\omega(k_p)}\|\Psi\|_\hhh+|\vp(k_p)|\||x| 
\Psi\|_\hhh\rk.$$
Since $\|\vp\|<\infty$, $\|\sqrt\omega\vp\|<\infty$ 
 and $\|\vp/\omega\|<\infty$, \kak{200} follows for 
$\Psi\in\EE$. By a limiting argument it can be extended for $\Psi\in\d$.  
\kak{201}  is rather easier than \kak{200}. 
We have for $\Psi\in{\cal C}$, 
$$\|\HP \HHH(\k_p,\k_q)\Psi\|_\hhh\leq \frac{e^2}{2}\sqrt{\omega(k_p)\omega(k_q)}|\vp(k_q)\vp(k_p)|
\||x|^2\Psi\|_\hhh.$$
Thus the lemma follows from a limiting argument and 
$\|\sqrt\omega\vp\|<\infty$.
\qed

From Lemma \ref{dia} the next lemma  immediately follows.

\bl{sin1}
For almost every $(\kn)\in \RR^{3n}$ it follows that 
$$\gr\in D(\kj b)\bigcap 
[\cap_p D(b(\k_1)...\widehat{b(\k_p)}...b(\k_n)(|x|+1))]$$
\eq{cap}
\bigcap 
[\cap_{q<p}(b(\k_1)...\widehat{b(\k_q)}...\widehat{b(\k_p)}...b(\k_n)|x|^2)]
\en 
and 
$$\|\kj b\gr\|_\hhh\leq 
\sump \delta_1(k_p) \|b(\k_1)...\widehat{b(\k_p)}...b(\k_n)(|x|+1) \gr\|_\hhh$$
$$+
\sump \sum_{q<p}
\delta_2(k_p, k_q) \|b(\k_1)...\widehat{b(\k_q)}...\widehat{b(\k_p)}...b(\k_n)|x|^2\gr\|_\hhh.
$$
\el
\proof 
Note that for  $(\kn)\not\in\g$ and $(\kn)\not=(0,...,0)$, 
$$\kj b \gr=\HP(\kn) \ov{\S 1 }(\knn)\gr+\HP(\kn) \S 2(\knn) \gr.$$
Let $\Psi\in\EE$ and $f_j\in\ccccc$, $j=1,...,n$. 
It is obtained that 
$$
\ji  |\F| \| \HP(\kn)\ov{\S 1 }(\knn)\Psi\|_\hhh \dk $$
\eq{pl}
\leq 
 \sump \ji  |\F| \delta_1(k_p) \|b(\k_1)...\widehat{b(\k_p)}...b(\k_n)(|x|+1) \Psi\|_\hhh \dk.
\en 
Similarly we obtain that 
$$\ji  |\F| \|\HP(\kn) {\S 2 }(\kn) \Psi \|_\hhh \dk $$
\eq{pll}
\leq  \sump \ji  |\F| 
\sum_{q<p}
\delta_2(k_p, k_q) 
\|{b(\k_1)...\widehat{b(\k_q)}...\widehat{b(\k_p)}...b(\k_n)|x|^2} 
\Psi \|_\hhh \dk.
\en 
We choose a sequence $\{\Psi_m\}\subset {\cal C}$ such that 
$\Psi_m\rightarrow \gr$ and 
$(\hff^K+|x|^{2K}) \Psi_m\rightarrow (\hff^K+|x|^{2K})\gr$ 
strongly  as $m\rightarrow \infty$ for sufficiently large $K$. 
Note that 
$|x|^j\Psi_m\rightarrow |x|^j\Psi$ and 
$(\hff^n+|x|^{2n})|x|^j\Psi_m\rightarrow (\hff^n+|x|^{2n})|x|^j\Psi$ 
strongly as $m\rightarrow \infty$ for $j=1,2$, since $K$ is sufficiently large. 
By Lemma \ref{ae} there exists a subsequence 
$\{m'\}\subset \{m\}$ such that 
for almost every $(\kn) \in \RRR$,   \kak{cap} follows and 
\eq{q}\kj b \Psi_{m'}\rightarrow \kj b \gr,\en 
\eq{qq}b(\k_1)...\widehat{b(\k_p)}...b(\k_n)(|x|+1) \Psi_{m'} 
\rightarrow b(\k_1)...\widehat{b(\k_p)}...b(\k_n)(|x|+1) \gr,
\en 
and 
\eq{qqq} 
b(\k_1)...\widehat{b(\k_q)}...\widehat{b(\k_p)}...b(\k_n)|x|^2\Psi_{m'} 
\rightarrow 
b(\k_1)...\widehat{b(\k_q)}...\widehat{b(\k_p)}...b(\k_n)|x|^2
\gr\en 
strongly as $m'\rightarrow \infty$. 
Moreover 
$$
 \ji |\F \delta_1(k_p)| 
 \|b(\k_1)...\widehat{b(\k_p)}...b(\k_n)(|x|+1) \Psid \|_\hhh\dk$$
$$
\leq 
\lk 
\int \delta_1(k_p)|f_p(k_p)|dk_p\rk 
\e'(f_1,...,\widehat{f}_p,...,f_n)\|(\hff^{n-1}+|x|^{2n-2})(|x|+1)\Psid\|_\hhh$$
and 
$$
\ji |\F  \delta_2(k_p, k_q)| 
 \|b(\k_1)...\widehat{b(\k_q)}...\widehat{b(\k_p)}...b(\k_n)|x|^2\Psid 
\|_\hhh\dk$$
$$
\leq \lk 
\int \delta_2(k_p,k_q))|f_p(k_p)f_q(k_q)|dk_pdk_q \rk$$
$$
\times \e'(f_1,...,\widehat{f}_p,...,\widehat{f}_q,...,f_n)
\|(\hff^{n-2}+|x|^{2n-4})|x|^2 \Psid\|_\hhh.$$
Then we have by \kak{pl}  and  \kak{pll} 
$$ 
\ji |\F| \|\kj b \Psi_{m'}\|_\hhh \dk $$
$$
\leq \sump \ji |\F \delta_1(k_p)| 
 \|b(\k_1)...\widehat{b(\k_p)}...b(\k_n)(|x|+1) \Psi_{m'} \|_\hhh\dk$$
$$+
\sump \sum_{q<p}
\ji |\F  \delta_2(k_p, k_q)| 
 \|b(\k_1)...\widehat{b(\k_q)}...\widehat{b(\k_p)}...b(\k_n)|x|^2\Psi_{m'} 
\|_\hhh\dk$$
$$\leq C \|(\hff^K+|x|^{2K})\Psi_{m'}\|_\hhh 
\leq C' \|(\hff^K+|x|^{2K})\gr\|_\hhh $$
with some constant $C$ and $C'$. 
Thus by the Lebesgue dominated convergence theorem and \kak{q}, \kak{qq} and \kak{qqq}, 
 we have 
$$ 
\ji |\F| \|\kj b \gr \|_\hhh \dk $$
$$
\leq \sump \ji |\F \delta_1(k_p)| 
 \|b(\k_1)...\widehat{b(\k_p)}...b(\k_n)(|x|+1) \gr \|_\hhh\dk$$
$$+
\sump \sum_{q<p}
\ji |\F  \delta_2(k_p, k_q)| 
 \|b(\k_1)...\widehat{b(\k_q)}...\widehat{b(\k_p)}...b(\k_n)|x|^2\gr
\|_\hhh\dk$$
Since $f_j\in\ccccc$, $j=1,...,n$, are arbitrary, the lemma follows. 
\qed

\bl{123}
Let $\Psi\in\d$. Then 
for almost every $(\kn)\in\RRR$ it follows that 
$$
\Psi\in D(\kj b)\bigcap 
[\cap_{l=0}^n \cap_{\{p_1,..,p_l\}\subset\{1,..,n\}} D(\kjj a |x|^l)]$$ and 
$$
\|\kj b\Psi\|_\hhh$$
\eq{R}
\leq 
\xxx \prod_{j=1}^l \frac{|e\vp(k_{p_j})|}{\sqrt{2\omega(k_{p_j})}}
\|\kjj a |x|^l\Psi\|_\hhh.\en 
\el
\proof 
Take a sequence $\{\Psi_m\}\subset \EE$ such that 
$\Psi_m\rightarrow \Psi$ and 
$(\hf^K+|x|^{2K}+1)\Psi_m\rightarrow  
(\hf^K+|x|^{2K}+1)\Psi$ strongly as $m\rightarrow \infty$ for sufficiently large $K$. 
\kak{R} is valid for  $\Psi$ replaced by $\Psi_m$, since 
$$\kj b\Psi_m=
(a(\k_1)+\t_1) ...  (a(\k_n)+\t_n)\Psi_m$$
$$=
\xxx \tl \kjj a \Psi_m .$$
Note that $|x|^l\Psi_m\rightarrow |x|^l\Psi$ and $(\hff^n+|x|^{2n})|x|^l\Psi_m\rightarrow 
(\hff^n+|x|^{2n})|x|^l\Psi$ 
 strongly as $m\rightarrow \infty$, since $K$ is sufficiently large. 
By Lemmas \ref{xa} and \ref{ae} there exists  
a subsequence $\mdd\subset\md$
such that for almost every $(\kn)\in\RRR$, 
$$\kj b\Psi_{m'}\rightarrow \kj b \Psi$$
and 
$$\kjj a |x|^l\Psi_{m'}\rightarrow \kjj a |x|^l\Psi.$$
Thus the proof is complete. 
\qed

\bl{321}
Let $\Psi\in\d$. Then for almost every $(\kn)\in\RRR$, 
$$
\Psi\in D(\kj a)\bigcap 
[\cap_{l=0}^n \cap_{\{p_1,..,p_l\}\subset\{1,..,n\}} 
D(\kjj b |x|^l)]$$ and 
$$\|\kj a\Psi\|$$ 
$$\leq 
\xxx \prod_{j=1}^l \frac{|e\vp(k_{p_j})|}{\sqrt{2\omega(k_{p_j})}}
\|\kjj b |x|^l\Psi\|_\hhh.$$
\el
\proof 
Note 
$\kj b =(a(\k_1)-\t_1)...(a(\k_n)-\t_n).$
The lemma is proven in the similar way as Lemma \ref{123}. 
\qed

\bl{l1}
Suppose that $|x|^z\Psi\in D(N^{n/2})\cap \d $ for $z=m,m+1,...,m+n$.  
Then there exist constants $\clk$ such that 
\eq{rnm}
\ji \|\kj b |x|^m \Psi \|_\hhh^2 \dk
\leq 
\YYY.
\en 
\el
\proof 
We have by Lemma \ref{123} and 
 $|\sum_{j=1}^N x_j|^2\leq N\sum_{j=1}^N x_j^2$,  
$$\ji \|\kj b |x|^m\Psi \|_\hhh^2 \dk
\leq 2^n \xxx \PPP ^l \times$$
$$\times\ji
\|\kjj a |x|^{m+l}\Psi \|_\hhh^2 \dkk.$$
By the assumption it  follows that $|x|^{m+l}\Psi \in D(N^{n/2})$. 
Thus we see that 
$$\ji\|\kjj a |x|^{m+l}\Psi \|_\hhh^2 \dkk$$
$$=\|\prod_{j=1}^{n-l}(N-j+1)^\han |x|^{m+l}\Psi \|_\hhh^2
\leq \XXX$$
with some constants $\alk$.
Then 
$$\ji \|\kj b |x|^m \Psi \|_\hhh^2 \dk$$
$$\leq \xxx \PPP^l\XXX.$$
Hence we conclude \kak{rnm}. 
\qed

We set the right hand side of \kak{rnm} by $\R_{n,m}(\Psi)$, i.e., 
$$\R_{n,m}(\Psi)=
\YYY.$$

\bl{l2}
Let $\Psi\in\d$.  Then there exist constants $\dln$ such that 
$$\ji  \|\kj a \Psi\|_\hhh^2 \dk$$
$$\leq 
2^n \lkk\ji \|\kj b \Psi \|_\hhh^2 \dk+
\ZZZ    \R_{n-l,l}(\Psi)\rkk.$$
\el
\proof
We have by Lemma \ref{321}, 
$$\ji \|\kj a\Psi\|_\hhh^2\dk
\leq 2^n \xxx \PPP^l \times$$
\eq{ppp}\times\int \|\kjj b|x|^l\Psi\|_\hhh^2 \dkk.
\en 
The term with $l=0$ in \kak{ppp} is just $\displaystyle \ji \|\kj b\Psi\|_\hhh^2 \dk$. 
The lemma follows from Lemma \ref{l1}.
\qed

{\it Proof of Theorem \ref{t1}}

We prove the theorem by means of an induction. 
It is known that 
$$\gr\in D(N^\han).$$ 
Suppose  that 
$$\gr\in D(N^{(n-1)/2}).$$ 
Then by Lemma \ref{b3}, 
\eq{tog}
\ji \|\kl a\gr\|^2_\hhh \dk<\infty,\ \ \ l=1,2,...,n-1,
\en 
and   by Lemma \ref{ess2}, 
\eq{fa}
\| N^{l/2} |x|^m \gr\|_\hhh  <\infty 
\en 
follows for all $m\geq 0$ and $l\leq n-1$. 
By Lemma \ref{l2} 
$$\ji \|\kj a\gr\|_\hhh^2\dk$$
$$\leq 2^n\lkk \ji\|\kj b\gr\|_\hhh^2\dk+
\ZZZ   \R_{n-l,l}(\Psi)\rkk.$$
By \kak{fa} 
we see that 
$$\R_{n-l,l}(\Psi)<\infty. $$ 
From Lemma \ref{sin1}   
it follows that 
$$\ji  \|\kj b  \gr\|_\hhh^2 \dk $$
$$\leq 
\delta_1 \sump  \ji  \|b(\k_1) ...  \widehat{b(\k_p)} ...  
b(\k_n) (|x|+1) \gr\|_\hhh^2 dk_1..\widehat{dk_p}
..  dk_n$$
\eq{w}
+
\delta_2 \sump \sum_{q<p} 
\ji  \|b(\k_1) ...  \widehat{b(\k_q)}  ...  \widehat{b(\k_p)} ...  b(\k_n)|x|^2\gr\|_\hhh^2
dk_1..\widehat{dk_q}.. \widehat{dk_p}
..  dk_n,
\en 
where 
$\delta_1=\int \delta_1(k)^2 dk$ and  $\delta_2=\int \delta_2(k,k')^2 dkdk'$. 
Then the right hand side of \kak{w} is finite by Lemma \ref{l1}. 
Hence 
$$\ji \|\kj a\gr\|_\hhh^2 \dk<\infty$$
follows, which implies, together with \kak{tog}, 
 that 
$$\gr\in D(N^{n/2})$$ by Lemma \ref{b3}. 
Thus the theorem follows.
\qed

{\it Proof of Theorem \ref{t2}}

This follows from Theorem \ref{t1} and Lemma \ref{ess}. 
\qed

\section{Appendix}
\subsection{Appendix A}
\bl{20}
Let $\Psi\in D(\hf^{n/2})$. 
Then there exists ${\cal M}(\Psi)\subset\RRR$ with the Lebesgue measure zero 
such that 
\eq{a1}
\Psi\in D(\kj a)
\en 
for $(\kn)\not\in{\cal M}(\Psi)$. 
Moreover assume that $\{\Psi_m\}\subset \EE$ satisfies that 
$\Psi_m\rightarrow \Psi$ and $\hff^{n/2}\Psi_m\rightarrow \hff^{n/2}\Psi$ 
strongly as $m\rightarrow \infty$. 
Then there exists a subsequence $\mdd\subset\md $ and 
 ${\cal M}(\Psi, \Psim, \mdd)\subset \RR^{3n}$ 
with the Lebesgue measure zero 
such that \kak{a1} follows and 
$$\lim_{m'\rightarrow \infty} \kj a \Psi_{m'}=\kj a \Psi$$
for $(\kn)\not\in {\cal M}(\Psi, \Psim ,\mdd)$. 
\el

\proof
We fix a sequence $\Psim$. 
The lemma is proven inductively. 
Note that 
\eq{40}
\|\hff^p\Psi\|\leq\|\hff^q\Psi\|\en for $p\leq q$. 
By \kak{o} we see that 
\eq {gg}
\ji  |f_1(k_1)|  \|a(\k_1) \Psi_m\|_\hhh 
dk_1\leq \e(f_1)\|(\hf+1)^\han\Psi_m\|_\hhh 
\en 
for arbitrary $f_1\in\ccccc$. 
The right hand side of \kak{gg} converges as $m\rightarrow \infty$ by \kak{40}. 
Then  the left hand side of \kak{gg} is a Cauchy sequence.  
Then there exist $\M 1\subset \BR$ with the Lebesgue measure zero 
 and a subsequence $\{m_1\}\subset \{m\}$ such that 
$a(\k_1)\Psi_{m_1}$ converges strongly as $m_1\rightarrow \infty$  for $k_1\not\in \M 1$.  
Since $a(\k_1)$ is closed, it follows that 
 for $k_1\not \in \M 1$, 
$\Psi\in D(a(\k_1))$ 
and 
$$s-\!\!\lim_{m_1\rightarrow \infty} a(\k_1)\Psi_{m_1}=a(\k_1)\Psi.$$ 
For $\Psi_{m_1}$ we have by \kak{o}
$$\ji |f_1(k_1)f_2(k_2)|\|a(\k_1)a(\k_2)\Psi_{m_1}\|_\hhh 
dk_1dk_2\leq \e(f_1,f_2)\|(\hf+1)\Psi_{m_1}\|_\hhh $$
for arbitrary $f_1,f_2\in\ccccc$. 
Then we also see that there exist $\M 2\subset \BR\times \BR$ with the Lebesgue measure zero and a subsequence $\{m_2\}\subset \{m_1\}$ such that 
$a(\k_1)a(\k_2)\Psi_{m_2}$ converges strongly as $m_2\rightarrow \infty$ for 
$(k_1,k_2)\not \in \M 2$.  
Since $a(\k_2)\Psi_{m_2}\rightarrow a(\k_2)\Psi$ strongly  as 
$m_2\rightarrow \infty$ for $k_2\not \in \M 1$ and $a(\k_1)$ is closed, 
we see that for 
$(k_1,k_2)\not \in \M 2\cup [\BR\times \M 1]$,  
$a(\k_2)\Psi\in D(a(\k_1))$ 
and 
$$s-\!\!\lim_{m_2\rightarrow \infty} 
a(\k_1)a(\k_2)\Psi_{m_2}= a(\k_1)a(\k_2)\Psi.$$ 
Repeating this procedure we see that there  exist 
subsets  $\M j\subset \RR^{3j}$, 
$j=1,...,n$,   with the Lebesgue measure zero and subsequences 
$\{m_n\}\subset \{m_{n-1}\}\subset ...\subset\{m\}$ such that 
 for $(\kn)\not \in \M n$, 
$\kj a \Psi_{m_n}$ 
converges strongly as $m_n\rightarrow \infty$ and 
$
a(\k_2)\dots a(\k_n)\Psi_{m_n}\rightarrow a(\k_2)\dots a(\k_n)\Psi$ 
strongly as $m_n\rightarrow \infty$ for 
$(k_2,...,k_n)\not \in \M {n-1}\cup 
[\BR\times \M {n-2}]\cup \dots \cup [\RR^{3(n-2)}
\times \M 1]$. 
Let $${\cal M}(\Psi,\Psim,\mdd)=\M n \cup [\BR\times \M {n-1}]\cup \dots \cup [
\RR^{3(n-1)}\times \M 1].$$
Since $a(\k_1)$ is closed, we see  that  for $(\kn)\not \in{\cal M}_p(\Psi,\Psim,\mdd)$,
$$a(\k_2)\dots a(\k_n)\Psi\in D(a(\k_1))$$ and 
 $$s-\!\!\lim_{m_n\rightarrow\infty}\kj a \Psi_{m_n}=\kj a \Psi.$$ 
Thus the proof is complete. 
\qed

We define $\ad_A^l(B)$ by $\ad^0_A(B)=B$ and $\ad^l_A(B)=[A, \ad^{l-1}_A(B)]$. 
Note that on $\fffw$ 
$$
[\hf^p, \kj a ]=\sum_{l=1}^p
\lk\!\!\!
\begin{array}{c}p\\l\end{array}
\!\!\!\rk \ad_{\hf}^l(\kj a )\hf^{p-l}, 
$$ 
$$\ad_{\hf}^l(\kj a)=
\sum_{p_1=0}^l 
\sum_{p_2=0}^{l-p_1} 
\dots 
\sum_{p_n=0}^{\lll}
\G{l}{p_1}\G{l-p_1}{p_2}\dots\G{\lll}{p_n}$$
$$
\times \ad_{\hf}^{p_1}(a(\k_1))\ad_{\hf}^{p_2}(a(\k_2))\dots 
\ad_{\hf}^{p_n}(a(\k_n)),$$
and 
$$\ad_{\hf}^p (a(\k))=(-1)^p  \omega(k)^p  a(\k).$$
Hence we have 
$$
[\hf^p, \kj a ]=\sum_{l=1}^p
\lk\!\!\!
\begin{array}{c}p\\l\end{array}
\!\!\!\rk 
\sum_{p_1=0}^l 
\sum_{p_2=0}^{l-p_1} 
\dots 
\sum_{p_n=0}^{\lll}
\G{l}{p_1}\G{l-p_1}{p_2}\dots\G{\lll}{p_n}$$
\eq{PP}
\times (-1)^l \omega(k_1)^{p_1}\omega(k_2)^{p_2}...\omega(k_n)^{p_n}
\kj a .
\en

\bl{ine}
Let $\Psi\in \ch $.  Then there exists 
${\cal M}_\infty(\Psi)\subset \RR^{3n}$ with the Lebesgue measure zero 
such that 
for  
$(\kn)\not \in {\cal M}_\infty(\Psi)$, 
$$\Psi\in D(\kj  a)$$ and 
$$\kj a \Psi\in  \ch. $$
\el
\proof
Let $ \{\Psi_m\}\subset \EE$ be such that 
$\Psi_m\rightarrow \Psi$ 
and $\hff^q\Psi_m\rightarrow \hff^q \Psi$ strongly  
as $m\rightarrow \infty$ for $q=(n/2)+p$. 
In particular 
$\hff^{n/2}\Psi_m\rightarrow \hff^{n/2}\Psi$ strongly as $m\rightarrow \infty$. 
By Lemma \ref{20}, 
 there exists a subsequence $\mdd\subset \md$ such that 
for $(\kn)\not \in {\cal M}(\Psi,\Psim,\mdd)$,  
$$\Psi\in D(\kj a )$$
and 
\eq{l}
\lim_{m'\rightarrow \infty}   \kj a \Psi_{m'} =\kj a \Psi.
\en 
We reset $m'$ as $m$. 
By \kak{PP}, for $f_j\in \ccccc$, $j=1,...,n$, 
$$\ji  |\F| \|\hf^p \kj a  \Psi_m\|_\hhh \dk$$
$$
\leq 
\ji  |\F| \|\kj a  \hf^p  \Psi_m\|_\hhh dk$$
$$
+
 \sum_{l=1}^p 
\lk \!\!\!\begin{array}{c}p \\l\end{array}
\!\!\!\rk 
\sum_{p_1=0}^l 
\sum_{p_2=0}^{l-p_1} 
\dots 
\sum_{p_n=0}^{\lll}
\G{l}{p_1}\G{l-p_1}{p_2}\dots\G{\lll}{p_n}
$$
$$
\times \ji |\F\omega(k_j)^{p_j}|   \|\kj a  \hf^{n-l}   \Psi_m\|_\hhh \dk
$$
$$
\leq \e(f_1,...,f_n) \|(\hf+1)^{n/2}\hf^p\Psi_m\|_\hhh$$
$$+
 \sum_{l=1}^p 
\lk \!\!\!\begin{array}{c}p \\l\end{array}
\!\!\!\rk 
\sum_{p_1=0}^l 
\sum_{p_2=0}^{l-p_1} 
\dots 
\sum_{p_n=0}^{\lll}
\G{l}{p_1}\G{l-p_1}{p_2}\dots\G{\lll}{p_n}$$
$$
\times \e(\omega^{p_1}f_1,...,\omega^{p_n}f_n)\|(\hf+1)^{n/2}\hf^{(p-l)}
\Psi_m\|_\hhh
$$\eq{rh}
\leq C\|\hff^{(n/2)+p}\Psi_m\|_\hhh
 \en
with some constant $C$. 
The right hand side of \kak{rh} converges strongly as 
$m\rightarrow \infty$. 
Since $f_j\in\ccccc$, $j=1,...,n,$ are  arbitrary, 
there exist $\M p\subset \RR^{3n}$ with the Lebesgue measure zero 
and  a subsequence $\{m'\}\subset \{m\}$ such that 
$\hf ^p \kj a \Psi_{m'}$  strongly converges   
as $m'\rightarrow \infty$ 
for $(\kn)\not \in\M p$. 
Since $\hf^p$ is closed, we obtain by \kak{l}  
that 
$$\kj a \Psi\in D(\hf^p)$$ for $(\kn) \not \in 
\Omega_p =\M p \bigcup {\cal M}(\Psi,\Psim,\mdd)$. 
Define 
$${\cal M}_\infty(\Psi) 
=\bigcup_p \Omega_p.$$ 
Then it follows that 
$\kj a \Psi\in \ch$  for $(\kn) \not \in {\cal M}_\infty$. 
\qed
{\it Proof of Lemma \ref{xa}}

Let $\{\Psi_m\}\subset  \EE$ be such that 
$\Psi_m\rightarrow \Psi$ and $(\hff^{n/2}+|x|^{2p})\Psi_m\rightarrow 
(\hff^{n/2}+|x|^{2p})\Psi$ 
strongly as $m\rightarrow \infty$. 
From Lemma \ref{ine} it follows that for $(\kn)\not 
\in {\cal M}_\infty(\Psi)$,   
$$\kj a \Psi\in \ch$$ and from Lemma \ref{20}  
\eq{41}
s-\!\!\lim_{m'\rightarrow \infty} 
\kj a \Psi_{m'}=\kj a \Psi
\en 
with  some subsequence $\{m'\}$ for $(\kn)\not\in {\cal M}(\Psi,\Psim,\mdd)$. 
We reset $m'$ as $m$. 
Let $f_j\in\ccccc$, $j=1,...,n$. 
Since $[|x|^p, \kj a]\Psi_m=0$, we have 
$$\lk \ji  |\F| \||x|^p \kj a \Psi_m\|_\hhh \dk\rk^2 $$
$$\leq \e(f_1,...,f_n)^2 
 \||x|^{2p}\Psi \|_\hhh\|\hff\Psi_m\|_\hhh
\leq 
\e(f_1,...,f_n)^2  \|(\hff+|x|^{2p})\Psi_m\|_\hhh^2.$$
Since the right hand side converges as $m\rightarrow \infty$, 
there exist  $\M p'\subset \RR^{3n}$ with 
the Lebesgue measure zero and a subsequence $\{m'\}$ 
such that  
 $|x|^p \kj a\Psi_{m'}$ 
strongly converge as $m'\rightarrow \infty$ for  $(\kn)\in \M p'$. 
Since  $|x|^p$  is closed and by \kak{41},  
$$\kj a \Psi\in D(|x|^p)$$ follows for 
$(\kn) \not\in \Omega_p'=
\M p'\bigcup {\cal M}(\Psi,\Psim,\mdd)$.  
Then  for $(\kn )\not \in \cup_p\Omega_p'$, 
$$\kj a \Psi\in \cx. $$ 
Let 
$${\cal M}_{\d}(\Psi,\Psim,\mdd)={\cal M}_\infty(\Psi)\bigcup[
\cup_p\Omega_p'].$$
Then the lemma follows. 
\qed
 
{\it Proof of Lemma \ref{xxa}}

Applying  \kak{op} instead of \kak{o}, we can show the 
lemma in the similar way as Lemmas \ref{20}, \ref{ine} and \ref{xa}. 
\qed

\subsection{Appendix B}
In this section we prove Lemma \ref{ess}.
In \cite{h11}  we proved that 
$e^{-tH}$ maps $D(\N )$ into itself for the case when $V=0$. 
We extend this result for some nonzero potential $V$. 
We  see  that if  $\gr\in D(\N)$ then the identity  
\eq{id}
\N \gr=\eht \ee \N \gr+\ee [\N, \eht]\gr
\en 
is well defined. 
Using \kak{id}  we shall prove  that $\|\N\gr(x)\|_\fff$ decays exponentially. 
To see it 
we prepare some probabilistic notations.

It is known that there exist a probability space $(Q,\mu)$ and 
Gaussian random variables 
$(\phi(f),f\in\od\LRr)$ such that 
$$\int _Q \phi(f)\phi(g) \mu(d\phi)=
\half \mn \int_\BR  \lk \delta_{\mu\nu} -\frac{k_\mu k_\nu}{|k|^2}\rk 
\overline{\hat{f}_\mu(k)} \hat{g}_\nu (k) dk.$$
For a general $f\in \od\LR $, we set $\phi(f)=\phi(\Re f) + i\phi(\Im f).$  
It is also known that there exists a unitary operator 
implementing 
$1\cong \Omega$, 
$L^2(Q)\cong \fff$ 
 and 
$\phi(\oplus_{\nu=1}^3 
 \delta_{\mu\nu} \lambda(\cdot-x))\cong A_\mu(x),$  
where $\lambda$ is the inverse Fourier transform of 
$$\hat\lambda=\vp/\sqrt\omega.$$
The free Hamiltonian in $\qqq$ corresponding to   $\hf$ in $\fff$ is denoted by $\hhf$. 
To have a functional integral representation of $e^{-t\hhf}$ we go  through 
another probability space $(Q_0, \nu_0)$  and Gaussian random variables 
$(\phi_0(f), f\in \od\LRnr)$ such that 
$$\int _{Q_0} \phi_0(f)\phi_0(g) \nu_0(d\phi_0)=
\half \mn \int_{\RR^4} \lk \delta_{\mu\nu} -\frac{k_\mu k_\nu}{|k|^2}\rk 
\overline{\hat{f}_\mu(k,k_0)} \hat{g}_\nu(k,k_0) dk dk_0.$$
Here $\phi_0(f)$ is also extended to $f\in\od \LRn$ such as $\phi(f)$. 
Let $j_t:\LR\rightarrow\LRn$ be the isometry defined by 
$$\widehat{j_t f}(k,k_0)=\frac{e^{-itk_0}}{\sqrt\pi} 
\sqrt{{\omega(k)}/({\omega(k)^2+|k_0|^2}) }\hat{f}(k)$$
and $J_t: \qqq\rightarrow \qqqq$ by 
$$J_t\wick{\phi(f_1) ... \phi(f_n)}=\wick{\phi_0([\od j_t] f_1) ... \phi_0([\od j_t] f_n)},$$
$$J_t1=1.$$
Here $\wick{X}$ denotes the Wick power  of $X$ inductively defined by 
$$\wick{\phis(f)}=\phis(f),$$
$$\wick{\phis(f) \phis(f_1)...\phis(f_n)}
=\phis(f)\wick{\phis(f_1)...\phis(f_n)}$$
$$-
\sum_{j=1}^n (\phis(f_j),\phis(f))_{L^2(Q_\ast)} \wick{\phis(f_1)...
\widehat{\phis(f_j)}...\phis(f_n)}, $$
where $Q_\ast=Q, Q_0$ and $\phis=\phi, \phi_0$. 
Then $J_t$ can be extended to an isometry and 
$J_t^\ast J_s=e^{-|t-s|\hhf}$ follows  for $t,s\in\RR$.   
We identify 
$\hhh=\LR\otimes\fff$ with  $L^2(\BR;\qqq)$. 
Under this identification $\Psi\in\hhh$ can be regarded as $\qqq$-valued $L^2$-function 
on $\BR$, 
i.e., $\Psi(x)\in\qqq$ 
for almost every $x\in\BR$. 
In \cite[Lemma 4.9]{h11} and \cite{h4} we established that 
$$\lk \eht \Psi\rk(x)=\EEEq \lk \vvv \J \Psi(X_t)\rk$$
for almost every $x\in\BR$. 
Here 
$(X_t)_{t\geq 0}=(X_{1,t},X_{2,t},X_{3,t})_{t\geq 0}\in C([0,\infty);\BR)$ denotes an $\BR$-valued  continuous path,   
$\EEEq$ an $\qqq$-valued  expectation value with respect to 
the wiener measure $P_x$ on $C([0,\infty);\BR)$ 
with $P_x(X_0=x)=1$, and 
$$\J=\J(x,X_{\cdot}): \qqq\rightarrow\qqq$$ is given by  
$$\J=J_0^\ast e^{-ie\phi_0(K(x,X_{\cdot}))} J_t,$$
where $K(x,X_{\cdot})$ is a $\od \LRn$-valued stochastic integral defined by 
$$K=K(x,X_{\cdot})=\odd \int_0^t j_s\lambda(\cdot-X_s) dX_{\mu, s}.$$  
Let $N$ and $\n$ be the number operators in $\qqq$ and $\qqqq$, respectively. 
Note that 
$$J_t N=\n J_t$$   
on a dense domain. 
The expectation value with respect to $P_x$ is denoted by  $\EEE$. 
We show a fundamental inequality.   
\bl{444}
Let 
$\K=\K(x,X_\cdot)=\|K(x,X_\cdot)\|_{\od\LRn}.$
Then,  for all $m\geq 0$,  
\eq{44}
\EEE \lk \K^{2m} \rk \leq 
\frac{3(2m)!}{2^m} t^{m-1} \EEE \lk 
\int_0^t \|j_s \lambda(\cdot-X_s)
\|^{2m}_\LRn  ds \rk 
= \frac{3(2m)!}{2^m} t^m  \|\vp/\sqrt\omega\|^{2m}.
\en
In particular 
$\supx  \EEE \lk \K^{2m} \rk<\infty.$
\el
\proof See \cite[Theorem 4.6]{h11}. 
\qed
\bl{234}
For each $(\x)\in \BR\times C([0,\infty);\BR)$ and $\Psi\in D(\N)$, 
$$\| [\N, \J(\x)] \Psi\|_\qqq  \leq P_k(\K ) \|(N+1)^{k/2}\Psi\|_\qqq,$$
with some polynomial $P_k(\cdot) $. 
\el
\proof
Note that for each $(\x)$, 
$\J=\J(\x)$ maps $D(\N)$ into itself. 
We have 
$$[\N, \J]\Psi=
J_0^\ast e^{-ie\phi_0(K)}[e^{ie\phi_0(K)} \NN e^{-ie\phi_0(K)}-\NN]J_t\Psi$$
$$=
J_0^\ast e^{-ie\phi_0(K)}
\lkk 
\lk 
N_0-e \phi_0'(K)+\frac{e^2}{2} \K \rk^{k/2}
-\N_0\rkk J_t\Psi$$
$$
=-\J \N \Psi+
J_0^\ast e^{-ie\phi_0(K)}
\lkk 
\lk 
N_0-e \phi_0'(K)+\frac{e^2}{2}\K\rk^{k/2}\rkk 
J_t\Psi,$$
where 
$\phi_0'(K)=i[N_0,\phi_0(K)].$ 
We see that 
$$\| \J \N\Psi\|_\qqq\leq \|\N \Psi\|_\qqq.$$ 
Note that 
$$\|\phi_0(K)\Psi\|\leq \sqrt 2  \K  \|(N_0+1)^\han\Psi\|.$$
Then it is obtained that 
$$\|
\lk 
N_0-e \phi_0'(K)+\frac{e^2}{2} \K \rk ^k J_t \Psi\|_\qqq\leq 
R_k(\K ) \|(N+1)^k\Psi\|_\qqq$$
with some polynomial $R_k(\cdot)$. 
Then 
$$\|[\N,\J]\Psi\|_\qqq 
\leq R_k(\K ) \|(N+1)^{k/2}\Psi\|_\qqq+\|\N\Psi\|_\qqq$$
$$\leq (R_k(\K)+1)\|(N+1)^{k/2}\Psi\|_\qqq.$$
Thus the proof is complete. 
\qed
\bp{simon}
Let $1\leq p\leq \infty$ and $a\geq 0$. Then 
there exists a constant $c_p =c_p(a)$ such that 
\eq{sim}
\supx \left|\EEE \lk e^{-a\int_0^t V(X_s) ds} f(X_t)\rk\right| 
\leq c_p \|f\|_{L^p(\BR)}.
\en 
\ep
\proof 
See \cite[Theorem B.1.1]{si}. \qed
\bl{q2}
We see that  $\eht$ maps $D(\N)$ into itself. 
\el
\proof
Let $\Phi,\Psi\in D(\N)$. 
We have 
$$(\N\Phi, \eht\Psi)_\hhh = 
\int \lk (\N\Phi)(x), \EEEq\lk \vvv\J\Psi(X_t)\rk\rk_\qqq dx$$
$$=\int   \EEE \lkk \lk \N\Phi(x) , \J \Psi(X_t)\rk_\qqq  \vvv\rkk dx.$$
Then 
$$(\N\Phi, \eht\Psi)_\hhh =\int   \EEE\lkk \lk \Phi(x),  \J \N \Psi(X_t)\rk_\qqq 
\vvv\rkk dx$$
$$+ 
\int   \EEE\lkk \lk \Phi(x),  [\N, \J]  \Psi(X_t)\rk_\qqq \vvv\rkk dx.
$$
Hence  we have by Lemma \ref{234}  
\eq{33}
|(\N\Phi, \eht\Psi)_\hhh|\leq \int  \EEE \lk 
\vvv \|\Phi(x)\|_\qqq \|\N\Psi(X_t)\|_\qqq\rk dx
\en
\eq{34}
+
\int  \EEE \lk 
\pk \vvv \|\Phi(x)\|_\qqq \|(N+1)^{k/2}\Psi(X_t)\|_\qqq\rk dx.
\en 
The first term \kak{33} is estimated as 
$$\kak{33} =  \lk \|\Phi(\cdot)\|_\qqq, \ehtp \|\N\Psi(\cdot)\|_\qqq\rk_\LR\leq 
e^{-tE_{\rm p}}\|\Phi\|_\hhh\|\N \Psi\|_\hhh,$$
where $E_{\rm p}=\inf\sigma(H_{\rm p})$. 
The second term \kak{34}  is estimated as
$$\hspace{-4cm} 
\kak{34}\leq \int  \|\Phi(x)\|_\qqq \times$$
$$\times \lk \EEE \pk ^2 \vvvv\rk^\han 
\lk \EEE\|(N+1)^{k/2}\Psi(X_t)\|_\qqq^2\rk^\han dx$$
$$\hspace{-4cm} \leq \int  \|\Phi(x)\|_\qqq \lk \EEE\pk ^4\rk^{1/4}\times$$
$$\times \lk \EEE \vvvvv\rk^{1/4} 
\lk \EEE\|(N+1)^{k/2}\Psi(X_t)\|_\qqq^2\rk^\han dx.$$
By Lemma \ref{444} we have  
$$\theta=\sup_{x\in\BR} \lk \EEE \pk^4\rk^{1/4} < \infty,$$
and by \kak{sim}, 
$$\eta=\sup_{x\in\BR} \lk \EEE \vvvvv\rk^{1/4} <\infty.$$ 
Then we have 
$$\kak{34}\leq  \theta \eta 
\int  \|\Phi(x)\|_\qqq \lk \EEE\|(N+1)^{k/2}\Psi(X_t)\|_\qqq^2\rk^\han dx $$
$$\leq \theta\eta  
\lk \int  \|\Phi(x)\|^2_\qqq dx \rk^\han 
\lk \int  \EEE  \|(N+1)^{k/2}\Psi(X_t)\|_\qqq ^2 dx \rk^\han$$
$$  = \theta\eta 
\|\Phi\|_\hhh \|(N+1)^{k/2}\Psi\|_\hhh.$$
Thus we conclude that 
$$|(\N\Phi, \eht \Psi)_\hhh|\leq \|\Phi\|_\hhh\lk e^{-tE_{\rm p}} 
\|\N \Psi\|_\hhh +\theta\eta \|(N+1)^{k/2}\Phi\|_\hhh\rk.$$
This implies that $\eht \Psi\in D(\N)$.
\qed

\bl{q3}
Assume that $\gr\in D(\N)$. Then  $\sup_{x\in\BR} \|\N\gr(x)\|_\qqq<\infty$.
\el
\proof 
By Lemma \ref{q2} the identity 
$\N\gr=\ee \eht \N\gr+\ee [\N,\eht]\gr$ 
is well defined, and we obtained 
that 
$$\N\gr(x)=\ee \EEEq\lk 
\vvv\J\N\gr(X_t)\rk 
+\ee \EEEq\lk 
\vvv[\N,\J]\gr(X_t)\rk $$
for almost every $x\in\BR$. 
We see that by Lemma \ref{234} 
\eq{ssd}
\|\N\gr(x)\|_\qqq\leq \ee \EEE\lk \vvv \|\N\gr(X_t)\|_\qqq\rk  
\en
\eq{sd}
+
\ee \EEE \lk \vvv \pk \|(N+1)^{k/2}\gr(X_t)\|_\qqq\rk.
\en
By \kak{sim} it is obtained that 
\eq{09}
\sup_{x\in\BR} \kak{ssd} <\infty.
\en 
\kak{sd} is estimated as 
$$\kak{sd}\leq \lk \EEE\pk^2\rk^\han\lk \EEE\vvvv\|(N+1)^{k/2}\gr(X_t)\|^2_\qqq\rk^\han.$$
By \kak{sim} 
we yield that 
$$\supx \EEE\lk 
\vvvv\|(N+1)^{k/2}\gr(X_t)\|^2_\qqq\rk <\infty,$$
and by Lemma \ref{444},   
$\sup_{x\in\BR} \EEE\lk \pk^2\rk<\infty$. 
Hence 
\eq{099}
\supx \kak{sd}<\infty.
\en 
Thus the lemma follows from \kak{09} and \kak{099}.
\qed

{\it Proof of Lemma \ref{ess}}\\
It is enough to prove the lemma for sufficiently large $|x|$ by Lemma \ref{q3}. 
Set $\theta=\sup_{x\in\BR} \|(N+1)^{k/2}\gr(x)\|_\qqq<\infty$. 
We have by \kak{ssd} and \kak{sd} for almost every $x\in\BR$ 
$$\|\N \gr(x)\|_\qqq\leq  \EEE\lk \vvv  (1+\pk)\rk \ee \theta $$
$$\leq \lkk \EEE\lk  (1+\pk)^2\rk \rkk^\han \lk \EEE\vvvv\rk^\han\ee \theta.$$
By \kak{44} we have 
$$\EEE \lk (1+\pk)^2\rk\leq Q_k(t),$$
where $Q_k$ is some  polynomial of the same degree  as $P_k$. 
Then we have 
$$\|\N \gr(x)\|_\qqq\leq \theta Q_k(t)  \ee \EEE \lk \vvvv \rk.$$ 
Here $t$ is arbitrary. 
Take $t=t(x) =|x|^{1-m}$. Then by \cite{ca} we see that  
there exist positive constants $D$ and $\delta$ such that for sufficiently large $|x|$, 
$$e^{t(x) E} \EEE\lk e^{-2\int_0^{t(x)} V(X_s) ds}\rk \leq  
De^{-\delta|x|^{m+1}}.$$ 
In the case of $m\geq 1$ it is trivial that 
$Q_k(t(x))\leq \theta'$ 
with some constant $\theta'$ independent of $x$. 
Hence 
$$\|\N \gr(x)\|_\qqq\leq \theta\theta' D e^{-\delta|x|^{m+1}}$$ 
follows for sufficiently large $|x|$. 
Thus the  lemma follows for 
$m\geq 1$. In the case of $m=0$, 
we see that $\|\N \gr(x)\|_\qqq\leq \theta Q_k(|x|)  D e^{-\delta |x|},$  
and hence 
$$\|\N \gr(x)\|_\qqq\leq \theta   D' e^{-\delta' |x|}$$
follows for $\delta'<\delta$ with some constant $D'$  
for sufficiently large $|x|$. 
The lemma is complete. 
\qed

{\bf Acknowledgment}
I thank M. Griesemer for pointing out an error in the first manuscript. 
This work is in part supported by 
Grant-in-Aid 13740106 for Encouragement of Young 
Scientists from the Ministry of Education, Science, Sports, and Culture.


{\footnotesize 
\begin{thebibliography}{99}
\bibitem{a} 
A.  Arai,  Ground state of the massless Nelson model without infrared cutoff 
in a non-Fock representation,   {\it Rev.  Math.  Phys. } {\bf 13} (2001), 
1075--1094. 

\bibitem{ah}
A.  Arai and M.  Hirokawa,   
On the existence and uniqueness of  ground states of 
a generalized spin-boson model,  
{\it J.  ~Funct.  ~Anal.  } {\bf 151}  (1997),   455--503.   


\bibitem{ahh}
A.  Arai,   M.  Hirokawa,   and  F.  Hiroshima,   
On the absence of eigenvectors of Hamiltonians in a class of massless quantum field models without infrared cutoff,   
{\it J.   Funct.   Anal.  } {\bf 168}  (1999),  470--497.  


\bibitem{bfs2}
V.  Bach,   J.  Fr\"ohlich,   I.  M.  Sigal,   
Quantum electrodynamics of confined nonrelativistic particles,   
{\it Adv.   Math.  } {\bf 137}  (1998),   299--395.  



\bibitem{bfs3}
V.  Bach,   J.  Fr\"ohlich,   I.  M.  Sigal,   
Spectral analysis for systems of atoms and molecules coupled to the quantized radiation field,   {\it Commun.   Math.   Phys.  } {\bf 207}  (1999),   249--290. 
\bibitem{bhlms}
V.  Betz,   F.  Hiroshima,   J.  L\H orinczi,   R.  A.  Minlos and  H.  Spohn,   
Gibbs measure associated with particle-field system,   
{\it Rev. Math. Phys.},  {\bf 14} (2002), 173--198. 

\bibitem{ca}
R.  Carmona,   Pointwise bounds for Schr\"odinger operators,   
{\it Commun.   Math.   Phys.  } {\bf 62}  (1978),   97--106.   



\bibitem{fgs}
J.  Fr\"ohlich,  M.  Griesemer and B.  Schlein,  
Asymptotic electromagnetic fields in a mode of quantum-mechanical 
matter interacting with the quantum radiation field,  
{\it Adv. in Math.} {\bf 164} (2001), 349--398. 


\bibitem{ge}
C.  G\'erard,   
On the existence of ground states for massless Pauli-Fierz Hamiltonians,   
{\it Ann.  Henri  Poincar\'e} {\bf 1}  (2000),  443--459.  


\bibitem{gll}
M.  Griesemer,   E.  Lieb and M.  Loss,   Ground states in non-relativistic quantum electrodynamics,   
{\it Invent. Math.} {\bf 145} (2001), 557--595. 
\bibitem{gr}
L.  Gross,   The relativistic Polaron without cutoffs, 
{\it Commun. Math. Phys.  } {\bf 31}   (1973),   25--73.  



\bibitem{h4}
F. Hiroshima,  
Functional integral representation of a model in quantum electrodynamics, 
{\it Rev.   Math.   Phys.} {\bf  9} (1997), 489-530.  

\bibitem{h8} 
F.  Hiroshima,  
Ground states of a model in nonrelativistic quantum electrodynamics I,   
{\it J.   Math.   Phys.  } {\bf 40}  (1999),   6209--6222, 
II, {\it J.    Math.   Phys.  }  {\bf 41}   (2000),   661--674.   
\bibitem{h11}
F.  Hiroshima,  
Essential self-adjointness of translation-invariant quantum field models for 
arbitrary coupling constants,   
{\it Commun.   Math.   Phys.  } {\bf 211}   (2000),   585--613.   


\bibitem{h16}
F.  Hiroshima,  
Self-adjointness 
of the Pauli-Fierz Hamiltonian for arbitrary values of coupling constants,   
{\it Ann. Henri  Poincar\'e}, {\bf 3} (2002), 171--201. 

\bibitem{h20} 
F.  Hiroshima,  
Analysis of ground states of atoms interacting with  a quantized radiation fields, 
to be published in {\it Int. J. Mod .Phys. B}.

\bibitem{hisp1}
F.  Hiroshima and H.  Spohn,   
Enhanced binding through  coupling to a quantum field,   
{\it Ann. Henri  Poincar\'e} {\bf 2} (2001), 1159--1187. 


\bibitem{hisp2}
F.  Hiroshima and H.  Spohn,   
Ground state degeneracy of the Pauli-Fierz model with   spin,  
{\it Adv. Theor. Math. Phys.} {\bf 5} (2001), 1091--1104. 

\bibitem{hvv} 
C. Hainzl, V. Vougalter and S. A. Vugalter, Enhanced binding in non-relativistic QED, 
mp-arc 01-455,  preprint, 2001.







\bibitem{lms}
J.  L\H{o}rinczi,   R.   A.  Minlos and H.  Spohn,  
The infrared behaviour in Nelson's model of a 
quantum particle coupled to a massless scalar field,  
{\it Ann. Henri  Poincar\'e} {\bf 3} (2001), 1--28. 



\bibitem{lms2}
J.  L\H{o}rinczi,   R.   A.  Minlos and H.  Spohn,  
Infrared regular representation of the three dimensional massless Nelson model, 
{\it Lett. Math. Phys.}  {\bf 59} (2002), 189--198. 


\bibitem{ne} 
E.  Nelson,   Interaction of nonrelativistic particles with a quantized scalar field,   {\it J.   Math.   Phys.  }{\bf  5}  (1964),   1190--1197.   



\bibitem{si}
B.  Simon,   
 Schr\"odinger semigroups,  {\it Bull.  Amer.  Math.  Soc.  }{\bf 7}  (1982),  447--526.  
{\it J.   Funct.   Anal.  } {\bf 32}  (1979),   97--101.   


\bibitem{sl}
A. Sloan, 
The polaron without cutoffs in two space dimensions, {\it J. Math. Phys.} {\bf 15} 
(1974), 190--201. 


\bibitem{sp}
H.  Spohn,   
Ground state of quantum particle coupled to a scalar boson field,   
{\it  Lett.   Math.   Phys.  } {\bf 44}   (1998),   9--16.   

\bibitem{sp3}
H.  Spohn,   
Ground state(s) of the spin-boson Hamiltonian,   {\it Commun.   Math.   Phys.  } 
{\bf 123}    (1989),   277--304.  



\end {thebibliography}}
\end{document}